\documentclass[final,1p,times]{elsarticle}
\usepackage{amsmath,amssymb,amsfonts,amsthm,enumerate,multirow,mathtools}
\usepackage{color}
\usepackage{siunitx}
\usepackage[linesnumbered,ruled]{algorithm2e} % for algorithms
\usepackage{diagbox} % for table
\usepackage{cleveref}
\usepackage{placeins} % for floatbarrier
\usepackage[table]{xcolor}% http://ctan.org/pkg/xcolor
\usepackage{graphicx}
\usepackage{tikz}
\usepackage{makecell}


\usepackage{lineno}
\journal{International Journal of Solids and Structures}

\begin{document}

\begin{frontmatter}

\title{An essay on $d_0$ in neutron-based strain measurement techniques: equilibrium-based methods for non-destructive $d_0$ estimation}

\author[Newc]{C.M. Wensrich\corref{cor}}
\ead{christopher.wensrich@newcastle.edu.au}
\author[Newc]{G.N. Parkes}
\author[ANSTO]{V. Luzin}
\author[Newc]{B.K. Daly}
\author[ANSTO]{F. Salvemini}

\cortext[cor]{Corresponding author}

\address[Newc]{School of Engineering, The University of Newcastle, University Drive, Callaghan, NSW 2308, Australia}
\address[ANSTO]{Australian Centre for Neutron Scattering, Australian Nuclear Science and Technology Organisation (ANSTO), Kirrawee NSW 2232, Australia}

\begin{abstract}
Determination of the stress-free reference lattice spacing, $d_0$, is a central and often underestimated difficulty in diffraction-based strain measurement. Although commonly treated as a material constant, $d_0$ often varies with position and measurement direction as a result of composition, phase, texture and residual stress at the sub-bulk scale.  While direct measurement of $d_0$ from coupons is usually preferable, this is sometimes impractical or even impossible. This paper reviews the $d_0$ problem in general and develops a hierarchy of non-destructive alternatives based on mechanical equilibrium. This ranges from established techniques based on force-balance, to the application of boundary conditions and finally an approach based on point-wise equilibrium within a sample, implemented through eigenstrain analysis.  In all cases, examples are provided in the form of real samples including an additively manufactured Inconel cube, ancient Roman bronze medical probes and an ancient bronze dagger of purported Persian origin. These latter examples demonstrate that equilibrium can provide a practical physical constraint for estimating $d_0$ when destructive reference measurements are entirely inappropriate.
\end{abstract}


\end{frontmatter}

\section{Introduction}

As a first-order approximation, every presentation and paper on neutron diffraction based strain measurement includes the following equation for elastic strain;
\begin{equation}
    \epsilon=\frac{d-d_0}{d_0}
\end{equation}
along with a dutiful explanation that $d$ is a measured lattice spacing and $d_0$ is the corresponding `reference' spacing in stress-free material. In this description it could be implied that $d_0$ is a material constant, but this simplistic view masks a wealth of subtle complexity that keeps instrument scientists up at night (never mind the 3:00AM sample change). In a practical sense, $d_0$ is not really an intrinsic property of the material, nor is it necessarily constant.  Needless to say, it is not something we can look up in a book.  Determining an appropriate value, or values, or distribution for $d_0$ is a key non-trivial task in any experiment of this type.  

First of all, it is important to remember that $d$-spacings in diffraction experiments are not measured directly; rather they are implied through Bragg's law
\[
n\lambda=2d\sin{\theta},
\]
where depending on the instrument, the diffraction angle $\theta$ or wavelength $\lambda$ is fixed and the other is measured in order to calculate $d$.  The measurement of $\theta$ or $\lambda$ inevitably involves fitting Gaussian distributions to observed histograms and this carries a host of systematic errors related to the configuration of the instrument and instrumentation.  In the context of making strain measurements, precision matters more than accuracy; particularly when it comes to $d_0$ and its relationship to the other measurements. In other words, it should contain all of the same systematic errors as the measurements in a consistent way. Usually this is achieved through direct measurement, but this is not always possible or practical.

In this essay, we will discuss the finer details of determining $d_0$ from the perspective of making accurate and meaningful strain measurements.  Our aim is to provide a primer to be read as due diligence in the planning of experiments.  Obviously there are many other things to consider when planning an experiment of this type and there is already a wealth of resources on the basics (e.g. \cite{withers_methods_2007,kisi_applications_2012,noyan_residual_2013,gnaupel-herold_measurement_2005,fitzpatrick_analysis_2003,hauk_structural_1997}). In this context, we will only cover what is required to focus on understanding the `$d_0$ problem' in all its guises and discuss modern techniques that supplement traditional approaches. 

Much has already been said about this topic (e.g. \cite{rogante_stress-free_2000,withers_methods_2007,maekawa_stress-free_2012}). Almost two decades ago, Withers \emph{et al} \cite{withers_methods_2007} published their seminal paper on obtaining the `strain-free' lattice parameter for diffraction methods in general (including X-ray techniques).  In terms of neutron measurements, much of what they wrote is still very much current practice in the area.  However, current practice with neutron measurements now extends to utilising equilibrium in more creative ways; particularly in situations where plane-stress assumptions can be made.  This was highlighted by Withers \emph{et al} \cite{withers_methods_2007}, but only in the context of the well-known `$\sin{}^2\psi$ technique' for X-ray techniques and overall stress/force balances as a sanity check. The reality is that equilibrium assumptions can be used in many more ways than this and for neutrons as well as X-rays.

In this paper we review the fundamentals of the $d_0$ problem from a dedicated `neutron' perspective.  After a brief review of the mechanics involved and the context of the problem, we catalogue a range of experimental and theoretical approaches to measuring and/or calculating $d_0$ with examples presented as appropriate.  Our predominant focus is on contrasting established methods (e.g. direct measurement from coupons) with new techniques based on equilibrium assumptions. At times the style of the essay could be seen as `teaching', however the techniques we will discuss are certainly novel.

We begin with an overview of why $d_0$ is rarely a simple matter.

\section{What is $d_0$?}

In some sense, the simple version is true; $d_0$ is the stress-free lattice spacing for the sample.  But what does this mean?  To illustrate the ambiguity, consider the following central question;

\vspace{2ex}
\begin{center}
    \emph{Is there a version of your sample that can exist without any stress at any length scale anywhere within the sample?}
\end{center}
\vspace{2ex}

If your answer is yes (Hint: it is not), would the lattice spacing within this imagined sample be in any way related to the sample you measured or the measurements you made?  The issue revolves around the various length-scales of stress within polycrystalline materials and the effect they can have on observed $d$-spacing.  

\begin{figure}
    \centering
    \includegraphics[width=0.85\linewidth]{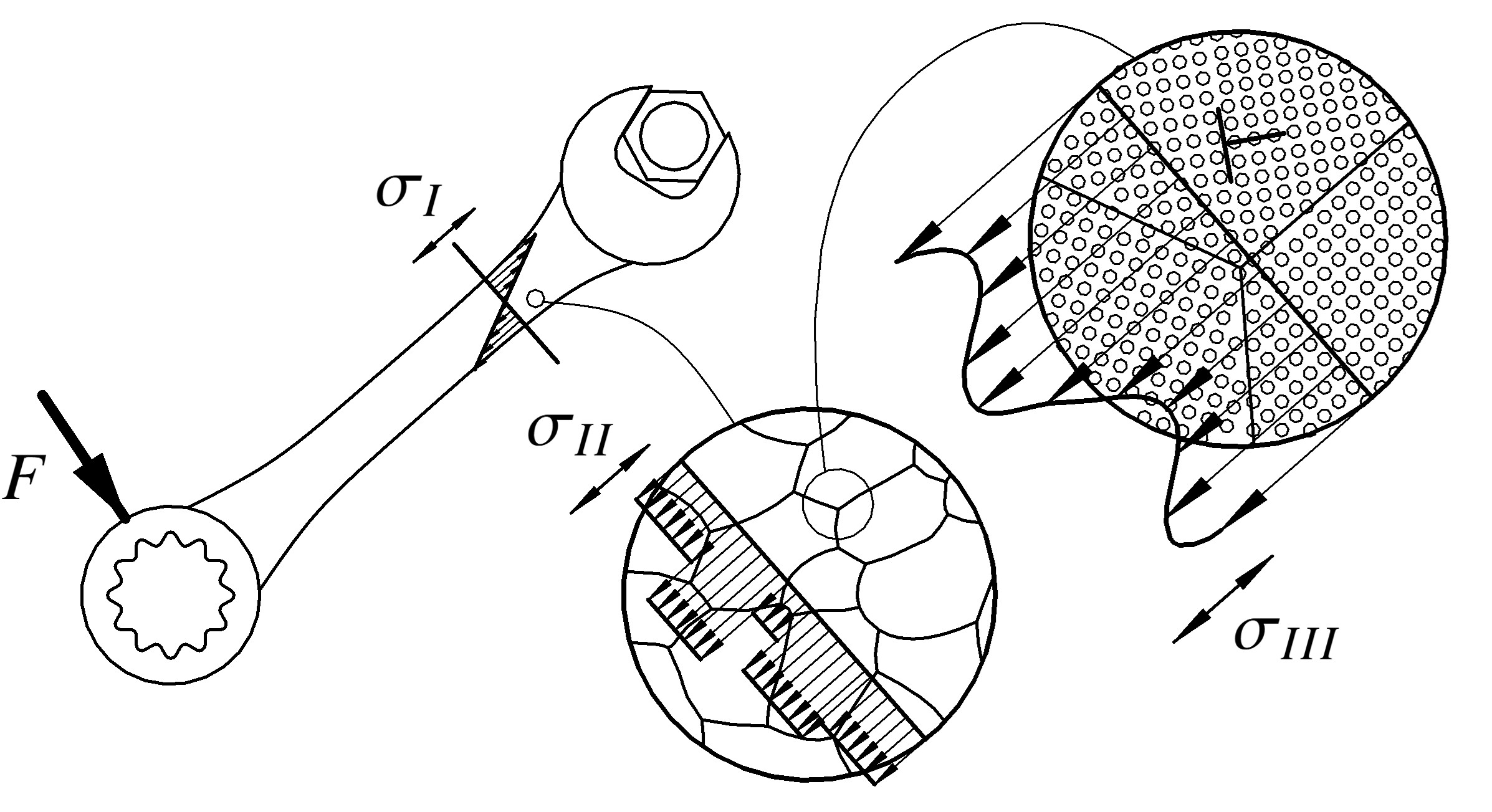}
    \caption{Stress in a polycrystalline object exists at various scales from the whole component down to the crystal lattice.}
    \label{fig:StressScales}
\end{figure}

As per \cite{withers_residual_2001} and Figure \ref{fig:StressScales}, three length scales are usually identified \cite{withers_residual_2001,noyan_residual_2013,kesavan_nair_residual_1995};
\begin{itemize}
    \item[--]{Type I: Macroscopic stress that equilibriates over the scale of the sample.}
    \item[--]{Type II: Inter-granular stresses that equilibriate between neighbouring grains, and,}
    \item[--]{Type III: Intra-granular stress that equilibriates within an individual grain (e.g. lattice stress/strain due to dislocations and inclusions/alloying elements)}
\end{itemize}

These form a decomposition where the stress at any given location is considered to be the sum
\[
\sigma=\sigma_I+\sigma_{II}+\sigma_{III}, 
\]
where over a large enough representative volume, the average of $\sigma_{II}$ and $\sigma_{III}$ are effectively zero.

The vast majority of neutron experiments are focused on measuring elastic strain through an observed $d$-spacing averaged over a gauge volume much larger than individual grains (this is not always the case with x-ray diffraction experiments). In this case, the measured average should depend entirely on $\sigma_I$, but the important question is: do $\sigma_{II}$ or $\sigma_{III}$ have any measurable impact? 

As a first-order approximation the answer is no; largely speaking, the strain associated with $\sigma_{II}$ and $\sigma_{III}$ averages to zero over the gauge volume and they only have an effect on the width of diffraction peaks rather than their position.  However, for a variety of reasons, this is not completely true.

In the case of $\sigma_{II}$, equilibrium is achieved through the action of neighbouring grains, but diffraction-based measurements do not `see' these neighbouring grains as it is unlikely that they align with the instrument geometry.  If the material has a non-uniform crystallographic texture and/or is anisotropic, we should consider the possibility that grains oriented in certain directions may, on average, experience slightly different $\sigma_{II}$ to the background.  This small deviation from `random' will manifest as a small change in the observed $d$-spacing.

Similarly for $\sigma_{III}$, equilibrium is achieved locally within the internal lattice structure of a grain, and our observation is restricted to a single set of lattice planes. Noting that single crystals are notoriously anisotropic; the effect of structured/aligned lattice dislocations (e.g. as a result of plastic strain and associated slip-systems) can cause a different level of strain in our particular set of lattice planes compared to the background.  Once again, this manifests as a small change in the observed $d$-spacing.

In both cases, if we are only interested in bulk stress (i.e. $\sigma_I$), the small shift in the observed $d$-spacing due to these effects should be considered part of our value for $d_0$.  Said another way; if we removed all forms of $\sigma_{II}$ and $\sigma_{III}$ stress from our sample, it would not be the same sample in terms of the reference state (i.e. with $\sigma_I=0$ alone).

The best case is that our sample contains all of these effects in a homogeneous and isotropic way.  In that case, we aim to measure (or calculate) a single $d_0$ as a reference for all of our spatially distributed measurements in all directions.  The worst case is that these effects are spatially distributed and directionally dependent; strictly speaking, $d_0$ defines its own deviatoric `reference' strain tensor field.

In addition to all of these stress-scale issues, there are also a host of other effects that can manifest as a `change' in $d_0$.  Catalogued by Withers \emph{et al} \cite{withers_methods_2007}, these include;
\begin{itemize}
    \item[--]{\emph{Composition Changes}; Tiny changes to the composition of a material can introduce a noticeable change to lattice spacing.  The example given by Withers \emph{et al} \cite{withers_methods_2007} (via \cite{bhadeshia_stress_1991}) is that the reference lattice spacing in steel can change by as much as $9\times 10 ^{-3}$ for each 0.1 wt\% of carbon in solution.  Many processes that we might be interested in (e.g. welding, sintering, additive manufacture, surface treatment) can generate uneven distribution of elements.  From this perspective, we are often confronted with the situation where the reference $d$-spacing varies significantly due to material composition.}
    \item[--]{\emph{Phase change}; Martensitic (and other) transformations are often accompanied by significant lattice distortion and, in many cases, appreciable volume changes. Like composition, there are a host of situations where the presence or extent of these transformations vary throughout a sample, and along with it the appropriate $d_0$.}
    \item[--]{\emph{Partial illumination and attenuation effects}; $d$-spacing measurements can be very sensitive to subtle shifts in the \emph{effective} centre of the gauge volume away from the geographic centre.  For measurements near boundaries, this can occur if part of the gauge volume lies outside the sample (noting that the edges of the gauge volume are not sharply defined in reality).  A similar effect can occur due to attenuation along varying paths through the gauge volume from the source to the detector.  Particularly in the case of large gauge volumes, these effects can manifest as significant changes to measured $d$-spacing well in excess of the peak fitting uncertainty.}
\end{itemize}

In this context, we will discuss various techniques for measuring, implying or calculating $d_0$ `in the wild'; constant or otherwise.  Various forms of these techniques can be found in the literature and we will present an overview and important considerations in each case. We will also introduce and develop a new approach capable of reconstructing $d_0$ distributions within samples based on point-wise application of the equilibrium constraint.   Before we begin, we start by reviewing the mechanics of elastic bodies, eigenstrain theory and how it relates to neutron-based strain measurement.

\section{Fundamentals of elasticity theory and residual stress}\label{Sec:Theory}

\begin{figure}
    \centering
    \includegraphics[width=0.25\linewidth]{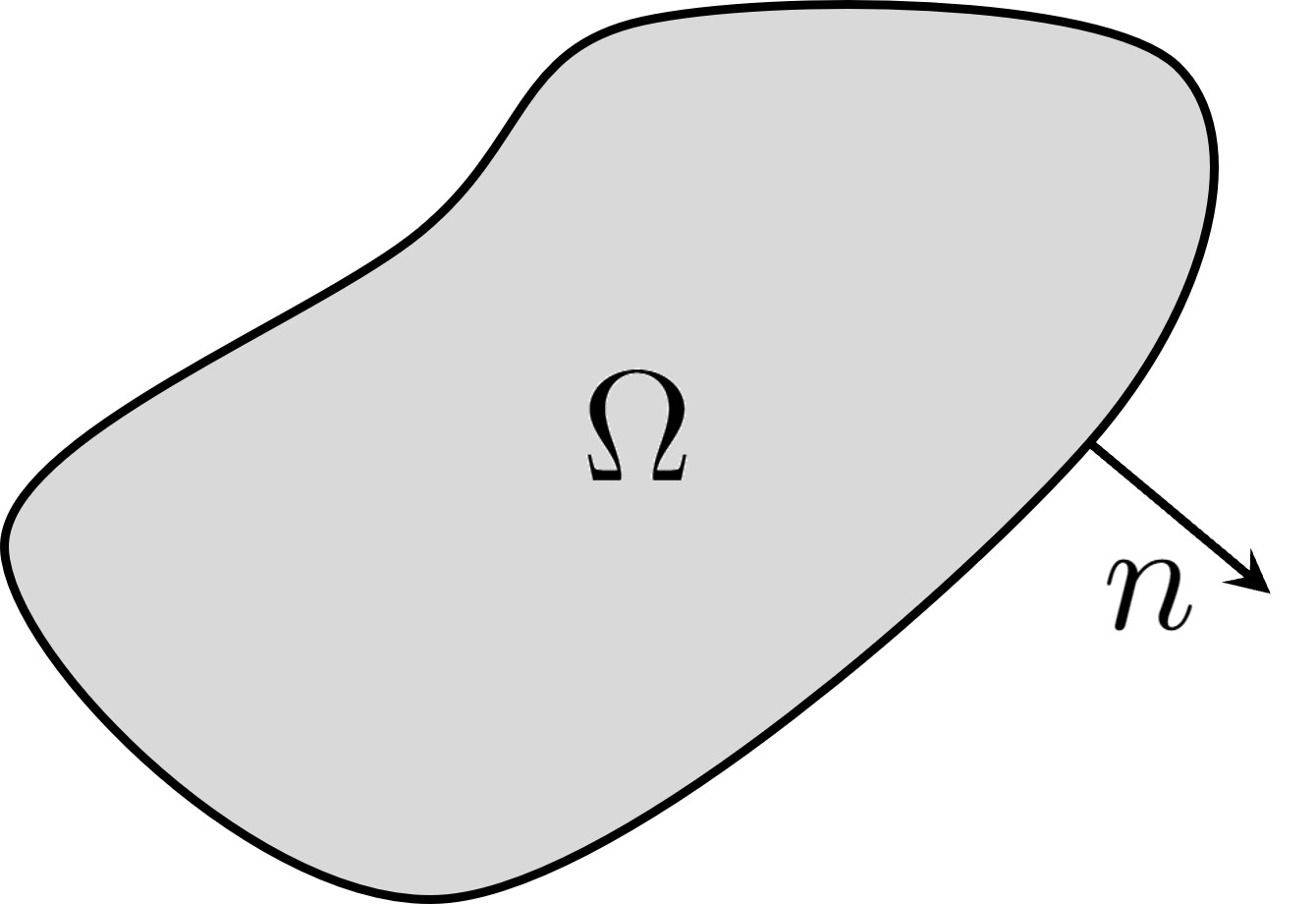}
    \caption{A finite sample $\Omega$ with unit normal $n$ on its surface $\partial\Omega$.}
    \label{fig:Omega}
\end{figure}

Consider a sample $\Omega$ with an outward unit normal vector $n$ on its smooth boundary $\partial\Omega$ (see Figure \ref{fig:Omega}).  Mechanical equilibrium implies
\begin{equation} \label{eq:Equilbrium}
\text{Div}(\sigma)=\nabla^T\cdot\sigma=\begin{bmatrix}
    \partial_1\sigma_{11}+\partial_2\sigma_{12}+\partial_3\sigma_{13}\\
    \partial_1\sigma_{12}+\partial_2\sigma_{22}+\partial_3\sigma_{23}\\
    \partial_1\sigma_{13}+\partial_2\sigma_{23}+\partial_3\sigma_{33}
\end{bmatrix}=
\begin{bmatrix}
    0\\
    0\\
    0
\end{bmatrix}
\end{equation}
at all points within $\Omega$.  Note that we have ignored gravity and other body forces that are typically negligible in neutron experiments.  Any applied loads are on the boundary and hence we can form the boundary condition
\begin{equation}\label{eq:BoundaryCond}
(\sigma \cdot n)|_{\partial\Omega}=\tau
\end{equation}
where $\tau$ is the (vector) concentration of applied traction force (i.e. stress) on the surface.  Practically, applied loads on samples are always distributed, but in this context we can consider any point load $F$ applied at a point $x_0\in\partial\Omega$ to be defined via the Dirac delta function
\[
\tau=F\delta(||x-x_0||).
\]
In the residual stress case we usually have $\tau=0$ everywhere on $\partial\Omega$.

We consider $\Omega$ to be a deformed version of itself with every point moved from some reference state by a small continuous displacement field $\chi(x)$.  The continuity of $\chi$ is important; otherwise we will have torn holes and folded material through itself to get to the deformed state.  From this displacement field we can define the total strain as the symmetric gradient
\begin{equation}\label{eq:StrainDisp}
\varepsilon=\nabla_s\chi=\tfrac{1}{2}(\nabla\chi^T+\chi\nabla^T)
\end{equation}
Part of this strain is elastic and is proportional to stress; the rest is inelastic and historic (e.g. plastic strain from ductile failure, phase change, misfits/interference, etc.) and is often termed `eigenstrain' \cite{mura_micromechanics_2013,korsunsky_eigenstrain_2008}.  From this we form the usual decomposition of total strain as
\begin{equation}\label{eq:EigenstrainDecomp}
\varepsilon=\epsilon+\epsilon^*,
\end{equation}
where $\epsilon^*$ is the inelastic eigenstrain and $\epsilon$ is the elastic strain related to stress through Hooke's law
\begin{equation}\label{eq:HookesLaw}
\epsilon=S:\sigma \quad \text{or} \quad \sigma = C:\epsilon,
\end{equation}
where $S$ and $C$ are rank-4 tensors of elastic constants and the $:$ operation implies a tensor inner product (i.e. contraction).  In the simplest case of a homogeneous isotropic material, all 81 coefficients of $S$ and $C$ can be written in terms of 2 parameters.  For example, in terms of Young's modulus $E$ and Poisson's ratio $\nu$, contraction with $S$ can be expressed
\[
\begin{split}
\epsilon_{11}&=\frac{1}{E}(+\sigma_{11}-\nu\sigma_{22}-\nu\sigma_{33})\\
\epsilon_{22}&=\frac{1}{E}(-\nu\sigma_{11}+\sigma_{22}-\nu\sigma_{33})\\
\epsilon_{33}&=\frac{1}{E}(-\nu\sigma_{11}-\nu\sigma_{22}+\sigma_{33})\\
\epsilon_{12}=\frac{1+\nu}{E}&\sigma_{12}, \enspace \epsilon_{13}=\frac{1+\nu}{E}\sigma_{13}, \enspace
\epsilon_{23}=\frac{1+\nu}{E}\sigma_{23}.
\end{split}
\]
A convenient isotropic version for $C$ that will be used in later calculations is
\[
\sigma=\frac{E}{(1+\nu)}\big(\epsilon+\alpha\text{tr}(\epsilon)I\big)
\]
where
\[
\alpha=\frac{\nu}{1-2\nu}
\]
and $I$ is the rank-2 identity.

Assuming $\tau=0$ on $\partial\Omega$, we can assemble \eqref{eq:Equilbrium}-\eqref{eq:HookesLaw} into the usual (linear) \emph{forward} eigenstrain model that maps a prescribed $\epsilon^*$ to the stress
\begin{equation}\label{eq:ForwardCalc}
\sigma=C:(\nabla_s\chi-\epsilon^*)
\end{equation}
where $\chi$ is the unique solution to the boundary value problem
\begin{equation}\label{eq:BVP}
\begin{split}
\mathrm{Div}(C:\nabla_s\chi)&=\mathrm{Div}(C:\epsilon^*) \quad \text{on} \quad \Omega,\\
(C:\nabla_s\chi)\cdot n&=(C:\epsilon^*)\cdot n \quad \text{on} \quad \partial\Omega.
\end{split}
\end{equation}
This boundary value problem may look daunting, however it is in the form of a typical structural elasticity problem where the right-hand-sides can be interpreted as a distributed body force and prescribed surface traction respectively.  Many commercial and open-source Finite Element packages exist that are specifically focused on solving this type of problem. 

We are also often interested in the \emph{inverse} problem of finding an eigenstrain that corresponds to a given distribution of stress.  In this respect there are several features of the eigenstrain mapping that are important to consider \cite{wensrich_well-posedness_2025,wensrich_uniqueness_2026,irschik_eigenstrain_2001,ziegler_eigenstrain_2004};
\begin{enumerate}
    \item{A large set of (non-zero) `nilpotent' or `null' eigenstrain fields exist that do not generate stress. This set consists of all compatible eigenstrains; i.e. any eigenstrain that can be written as $\epsilon^*=\nabla_s U$ for some underlying displacement field $U$ will not generate any residual stress.}
    \item{The existence of this null set means that we do not need to search over all possible eigenstrains to solve an inverse problem (many of them do nothing).  Without loss of generality, we can limit the search by either considering strictly diagonal eigenstrains (see \cite{wensrich_uniqueness_2026}) or divergence-free eigenstrains (see \cite{wensrich_well-posedness_2025}).  In either case the search can be significantly reduced by representing eigenstrain as three scalar fields instead of six independent components.}
    \item{Any inverse eigenstrain problem has a trivial solution of the form $\epsilon^*=-S:\sigma$ where in this case $\sigma$ is the known stress distribution we are trying to match \cite{wensrich_well-posedness_2025}.  Any solution to an inverse eigenstrain problem can be considered to be this trivial solution plus something from the null set.}
\end{enumerate}

From the perspective of the last of these observations we might wonder if there is any point to solving inverse eigenstrain problems.  In the direct sense, probably not; however the eigenstrain framework provides a powerful tool for generating arbitrary physically admissible strain fields that are useful for solving all manner of inverse problems related to stress and strain (e.g. \cite{korsunsky_eigenstrain_2008,korsunsky_teaching_2017,korsunsky_variational_2007,uzun_eigenstrain_2025,uzun_tomographic_2024,uzun_voxel-based_2023,wensrich_residual_2024,wensrich_well-posedness_2025}).

The key point to recognise is that we would like our strain measurements to only relate to $\epsilon$.  Indeed, plasticity within a material largely has no effect on $d$-spacing (aside from the generation of lattice dislocations).  However, the same cannot be said about other sources of eigenstrain such as phase change and thermal expansion.  In general, we consider variation in $d$-spacing due to eigenstrain as being part of our definition of $d_0$ which may no longer be constant throughout the sample.  In this context we will show later in the paper how the solution to inverse eigenstrain problems can be useful for indirectly determining a spatially varying $d_0$ due to these effects.

To build to this point, we will now discuss a variety of existing techniques used to determine $d_0$ in practice.

\section{Direct measurement of $d_0$}

The simplest and most common approach to determine $d_0$ is also the most obvious; relieve the stress and measure the $d$-spacing using an identical instrument setup.  However, in the context of the above it is important to recognise we are attempting to remove $\sigma_I$ and leave $\sigma_{II}$ and $\sigma_{III}$ intact.  It is also important that we preserve the composition and phase of material within the sample.  All of these conditions effectively rule out any process of stress relaxation through thermal/heat treatment.

Overwhelmingly, stress relaxation in this context is achieved through dissection; ideally cutting the sample into small stress-free coupons (e.g. cubes, combs, etc).  

Much has already been written about this approach (e.g. \cite{withers_methods_2007}) and we see no particular need to add anything further except to say the method of cutting is important.  This is in the context that we should aim to leave the material as close to its original condition/state - just relieve the stress, nothing more.  We should avoid the generation of excessive heat and plastic strain around the cut surface as much as possible.  Effectively all cutting processes introduce their own localised residual-stress fields around the cut surface, and it is important to recognise that any directly measured $d_0$ may include systematic errors of this kind.

\section{Estimating $d_0$ from force-balance and boundary conditions}

\begin{figure}
    \centering
    \includegraphics[width=0.4\linewidth]{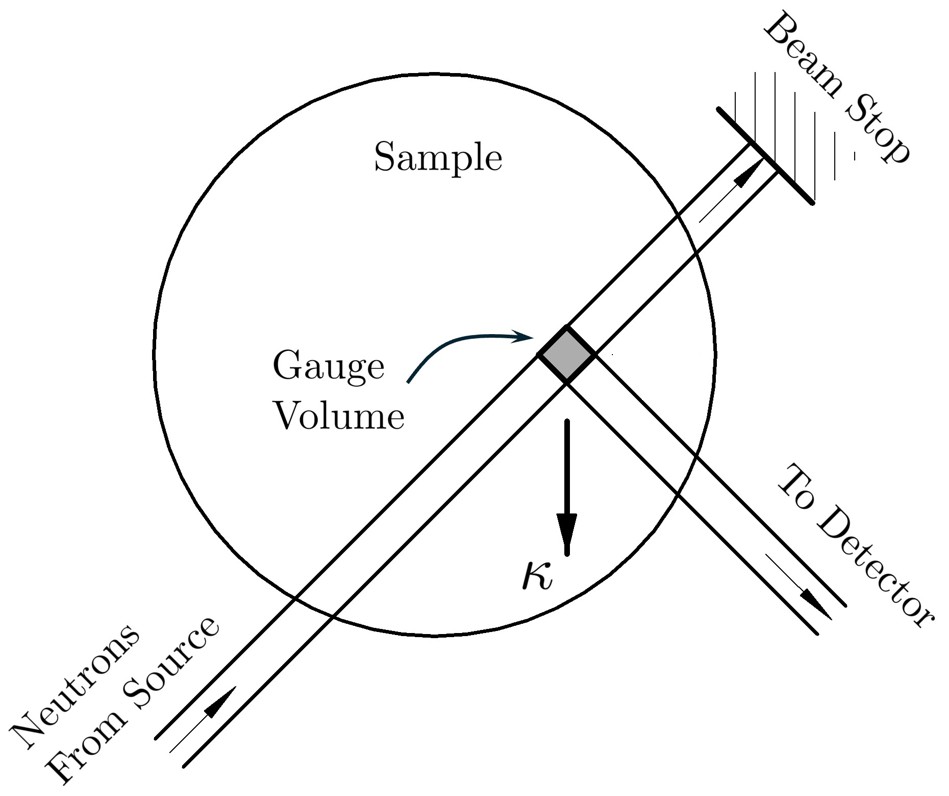}
    \caption{The geometry of a typical neutron-based strain measurement where a gauge volume is defined by masked incident and diffracted beams and average normal strain in this gauge volume is measured in the $\kappa$ direction.}
    \label{fig:DiffractionGeometry}
\end{figure}

Strain is a tensor and each measurement we make only probes a single component.  With reference to the typical measurement geometry shown in Figure \ref{fig:DiffractionGeometry}, the measured strain at a point is the average over the gauge volume of the quantity
\begin{equation}
\label{eq:StrainComponent}
\epsilon_{\kappa}=\kappa^T\epsilon\kappa=\epsilon_{11}\kappa_1^2+\epsilon_{22}\kappa_2^2+\epsilon_{33}\kappa_3^2+2\epsilon_{12}\kappa_1\kappa_2+2\epsilon_{13}\kappa_1\kappa_3+2\epsilon_{23}\kappa_2\kappa_3
\end{equation}
where $\kappa$ is the unit vector bisecting the incident and diffracted beams.

There are six unknown components of strain and so we need six independent measurements of this form to determine $\epsilon$ at any given point.  As an exercise for the reader, it is possible to show the necessary condition for the resulting system of equations to be independent is that the six unit vectors cannot lie on a common cone with its apex at the origin.  

But why stop at six? Every measurement we make reduces the uncertainty a small amount and with many measurement directions we can `over determine' the system as much as we like.  If we measure strain in $m$ directions $\{\kappa_{(1)}\cdots\kappa_{(m)}\}$, we can form the system
\[
[K]_{m\times6}[\epsilon]_{6\times1}=[\mathcal{E}]_{m\times 1}
\]
\[
\begin{bmatrix}
\kappa_{(1)1}^2 & \kappa_{(1)2}^2 &\kappa_{(1)3}^2 & 2\kappa_{(1)1}\kappa_{(1)2} & 2\kappa_{(1)1}\kappa_{(1)3} & 
2\kappa_{(1)2}\kappa_{(1)3}\\
\kappa_{(2)1}^2 & \kappa_{(2)2}^2 &\kappa_{(2)3}^2 & 2\kappa_{(2)1}\kappa_{(2)2} & 2\kappa_{(2)1}\kappa_{(2)3} & 
2\kappa_{(2)2}\kappa_{(2)3}\\
 & & & \cdots & & &\\
 \kappa_{(m)1}^2 & \kappa_{(m)2}^2 &\kappa_{(m)3}^2 & 2\kappa_{(m)1}\kappa_{(m)2} & 2\kappa_{(m)1}\kappa_{(m)3} & 
2\kappa_{(m)2}\kappa_{(m)3}
\end{bmatrix}
\begin{bmatrix}
\epsilon_{11}\\
\epsilon_{22}\\
\epsilon_{33}\\
\epsilon_{12}\\
\epsilon_{13}\\
\epsilon_{23}
\end{bmatrix}=
\begin{bmatrix}
    \epsilon_{\kappa_{(1)}}\\
    \epsilon_{\kappa_{(2)}}\\
    \cdots\\
    \epsilon_{\kappa_{(m)}}\\
\end{bmatrix}.
\]
From this, the `least squares' best fit to the data can be found using the Moore-Penrose pseudo-inverse $[K]^+=([K]^T[K])^{-1}[K]^T$, giving the strain at our point as
\begin{equation}\label{eq:LeastSquaresStrain}
[\epsilon]=[K]^+[\mathcal{E}].
\end{equation}

With so much data, perhaps we can do even more.  Writing \eqref{eq:StrainComponent} in terms of the measured $d$-spacings we have
\begin{equation}
\label{eq:d_kappa}
d_\kappa=d_0\kappa^T\epsilon\kappa +d_0,
\end{equation}
and from this we can write a similar system of equations, but now with $d_0$ included as an unknown to be determined from the data.  This provides
\[
[K]_{m\times7}[d_0\epsilon]_{7\times1}=[\mathcal{D}]_{m\times 1}
\]
\begin{equation}\label{eq:fitd0}
\begin{bmatrix}
\kappa_{(1)1}^2 & \kappa_{(1)2}^2 &\kappa_{(1)3}^2 & 2\kappa_{(1)1}\kappa_{(1)2} & 2\kappa_{(1)1}\kappa_{(1)3} & 
2\kappa_{(1)2}\kappa_{(1)3} &1\\
\kappa_{(2)1}^2 & \kappa_{(2)2}^2 &\kappa_{(2)3}^2 & 2\kappa_{(2)1}\kappa_{(2)2} & 2\kappa_{(2)1}\kappa_{(2)3} & 
2\kappa_{(2)2}\kappa_{(2)3} &1\\
 & & & \cdots & & &\\
 \kappa_{(m)1}^2 & \kappa_{(m)2}^2 &\kappa_{(m)3}^2 & 2\kappa_{(m)1}\kappa_{(m)2} & 2\kappa_{(m)1}\kappa_{(m)3} & 
2\kappa_{(m)2}\kappa_{(m)3} &1
\end{bmatrix}
\begin{bmatrix}
d_0\epsilon_{11}\\
d_0\epsilon_{22}\\
d_0\epsilon_{33}\\
d_0\epsilon_{12}\\
d_0\epsilon_{13}\\
d_0\epsilon_{23}\\
d_0
\end{bmatrix}=
\begin{bmatrix}
    d_{\kappa_{(1)}}\\
    d_{\kappa_{(2)}}\\
    \cdots\\
    d_{\kappa_{(m)}}\\
\end{bmatrix}.
\end{equation}
Unfortunately there is a problem with this strategy.  In each row, $\kappa_{(i)}$ is a unit vector with \[||\kappa_{(i)}||=\kappa_{(i)1}^2+\kappa_{(i)2}^2+\kappa_{(i)3}^2=1,
\]
and so the sum of the first 3 columns is always equal to the last column.  In other words, the rank of this matrix is at most 6 and we cannot uniquely solve for the unknowns.  In particular, null vectors exist of the form $[d_0\epsilon]^T=\alpha\begin{bmatrix}1&1&1&0&0&0&-1\end{bmatrix}$ which can be added to any possible solution without affecting the measured $d$-spacing.

This can be understood from a more physical standpoint by decomposing our strain into separate hydrostatic (i.e. isotropic) and trace-free deviatoric parts
\[
\epsilon=\epsilon^d+\bar\epsilon I,
\]
where $\bar\epsilon=\tfrac{1}{3}\text{tr}(\epsilon)$ and $\epsilon^d=\epsilon-\bar\epsilon I$. From \eqref{eq:d_kappa}, we can then write each measurement as
\begin{equation} \label{eq:decomposition}
d_\kappa=d_0(\kappa^T\epsilon^d\kappa) + d_0(1+\bar\epsilon),
\end{equation}
noting that $d_0(1+\bar\epsilon)$ is a constant term that does not depend on $\kappa$.  What if the deviatoric part was zero?  Every measurement from every direction would give us the same constant term and it would be impossible to separately determine both $d_0$ and $\bar\epsilon$ from this one constant.  Herein lies the essential `$d_0$ problem'; a set of diffraction measurements alone cannot distinguish between a change in the reference lattice spacing and a uniform hydrostatic strain. 

Interestingly, we do not need to know a precise $d_0$ to determine the deviatoric component.  Say we assign a value of $\tilde d_0$ as a reasonable guess.  From \eqref{eq:decomposition}, these measured $d$-spacings would then imply strains of
\begin{equation}\label{eq:DevHydDecomp}
\tilde\epsilon_\kappa=\frac{d_\kappa-\tilde d_0}{\tilde d_0}=\frac{d_0}{\tilde d_0}(\kappa^T\epsilon^d\kappa)+e
\end{equation}
where
\[
e=\frac{d_0-\tilde d_0}{\tilde d_0}+\frac{d_0}{\tilde d_0}\bar\epsilon
\]
is a constant that would be interpreted as the (incorrect) hydrostatic component of strain. The reconstructed deviatoric component of strain would be
\[
\tilde\epsilon^{d}=\frac{d_0}{\tilde d_0}\epsilon^d.
\]
Given that elastic strain is small, the measurements themselves give us an estimate for $d_0$ to within $0.5\%$ or better.  So this estimate gives $\frac{d_0}{\tilde d_0}\approx 1$ and $\tilde\epsilon^{d}\approx\epsilon^d$ to within a fraction of a percent.  Note that this is enough to compute a very good estimate of the von Mises stress at every measurement location in the form
\[
\sigma_{vM}=\frac{E}{2(1+\nu)}\sqrt{6\tilde \epsilon^d:\tilde\epsilon^d}.
\]

Returning to the actual $d_0$ problem, we have established that additional information is always required beyond the just the $d$-spacing measurements. In the direct approach discussed earlier, this additional information is that all components of strain are zero (i.e. in a specially constructed stress-free $d_0$ sample), but we do not always have to be so extreme.  Within our sample there are often boundary conditions that must be met and overall force balances that can be applied. We discuss each of these as follows;

\subsection{$d_0$ from force balance} \label{sec:ForceBal}

Consider the common scenario of a set of measurements made over a planar slice or cross-section within the sample $\Omega'$.  It is a basic principle that the stress on this cross-section must be in overall equilibrium with respect to the forces applied to the sample on either side of the slice.  In the case of residual stress this usually means the net force integrated over the cross-section is zero.  As observed above, $d_0$ has (almost) no bearing on measured shear stress, but the same is not true for normal stress and the overall stress-balance normal to the slice can be used to determine an accurate estimate for $d_0$.  

Inevitably the measurement geometry is usually aligned with the slice and we will simplify the analysis based on the assumption that $\Omega'$ is perpendicular to the $x_1$-axis.  In this case the stress-balance takes the form
\begin{equation}\label{eq:d0viaForceBal}
\begin{split}
0&=\int_{\Omega'}\sigma_{11} \rm{d}A\\
0&=\int_{\Omega'}(1-\nu)\epsilon_{11}+\nu(\epsilon_{22}+\epsilon_{33}) \rm{d}A\\
0&=\int_{\Omega'}(1-\nu)\frac{d_1-d_0}{d_0}+\nu\frac{d_2+d_3-2d_0}{d_0} \rm{d}A\\
d_0&=\frac{1}{(1+\nu)A_{\Omega'}}\int_{\Omega'}(1-\nu)d_1+\nu(d_2+d_3) \rm{d}A
\end{split}
\end{equation}
where $A_{\Omega'}$ is the area of the cross section, and $d_1$, $d_2$ and $d_3$ are the normal and two orthogonal in-plane observed $d$-spacings respectively.  All that remains is to numerically estimate the value of this integral from discrete measured data.

\begin{figure}
    \centering
    \includegraphics[width=0.32\linewidth]{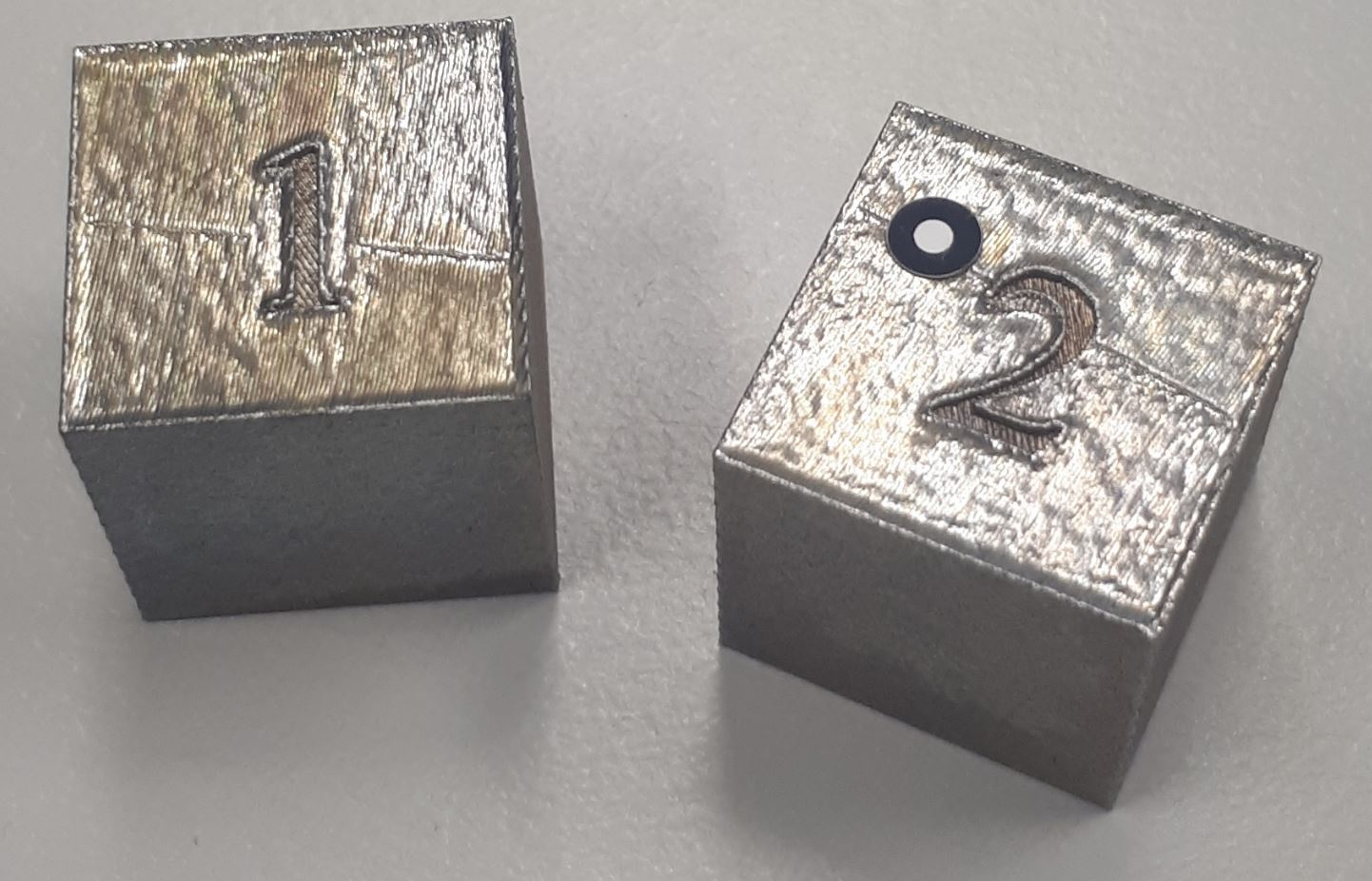}
    \includegraphics[width=0.38\linewidth]{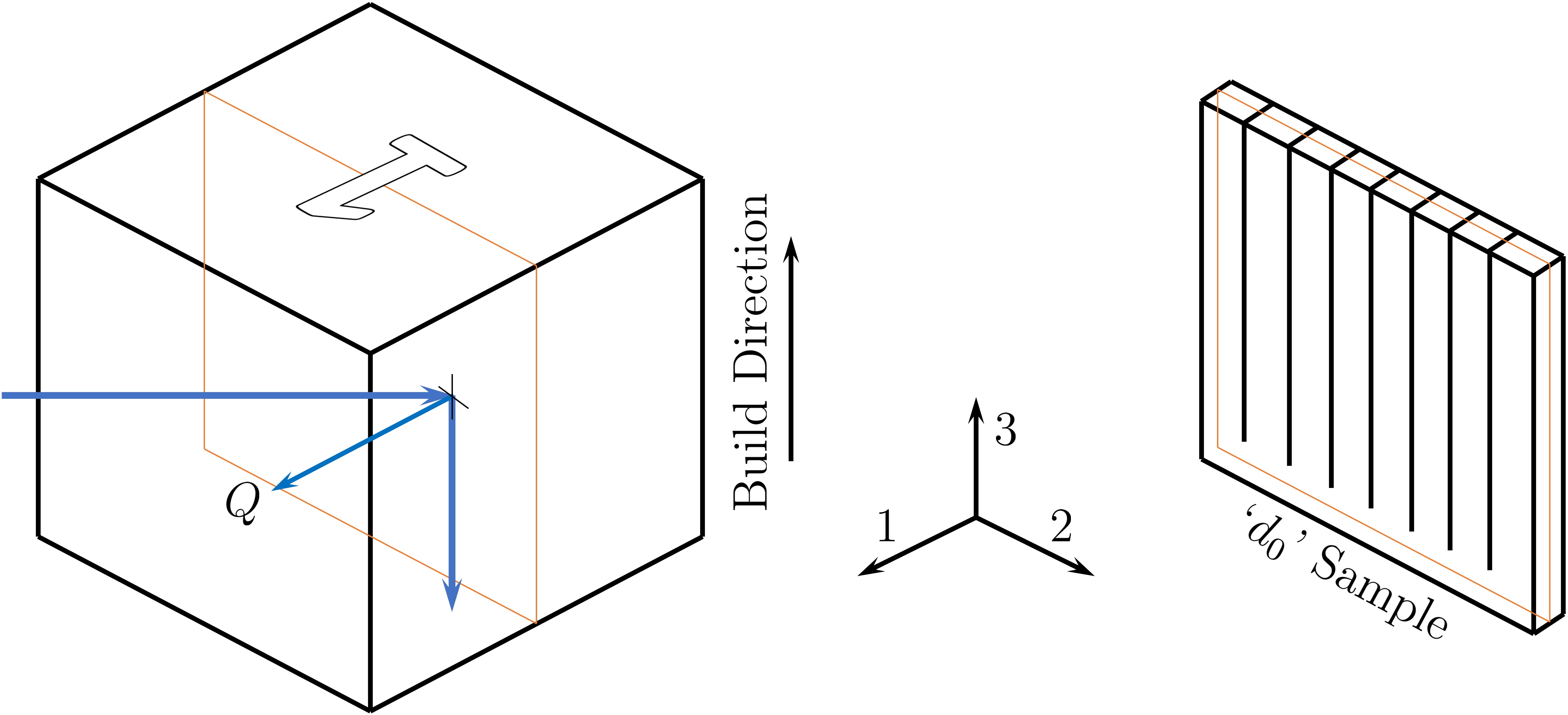}
    \put(-270,82){(a)}
    \put(-145,82){(b)}
    \hspace{0.5ex}
    \includegraphics[width=0.22\linewidth]{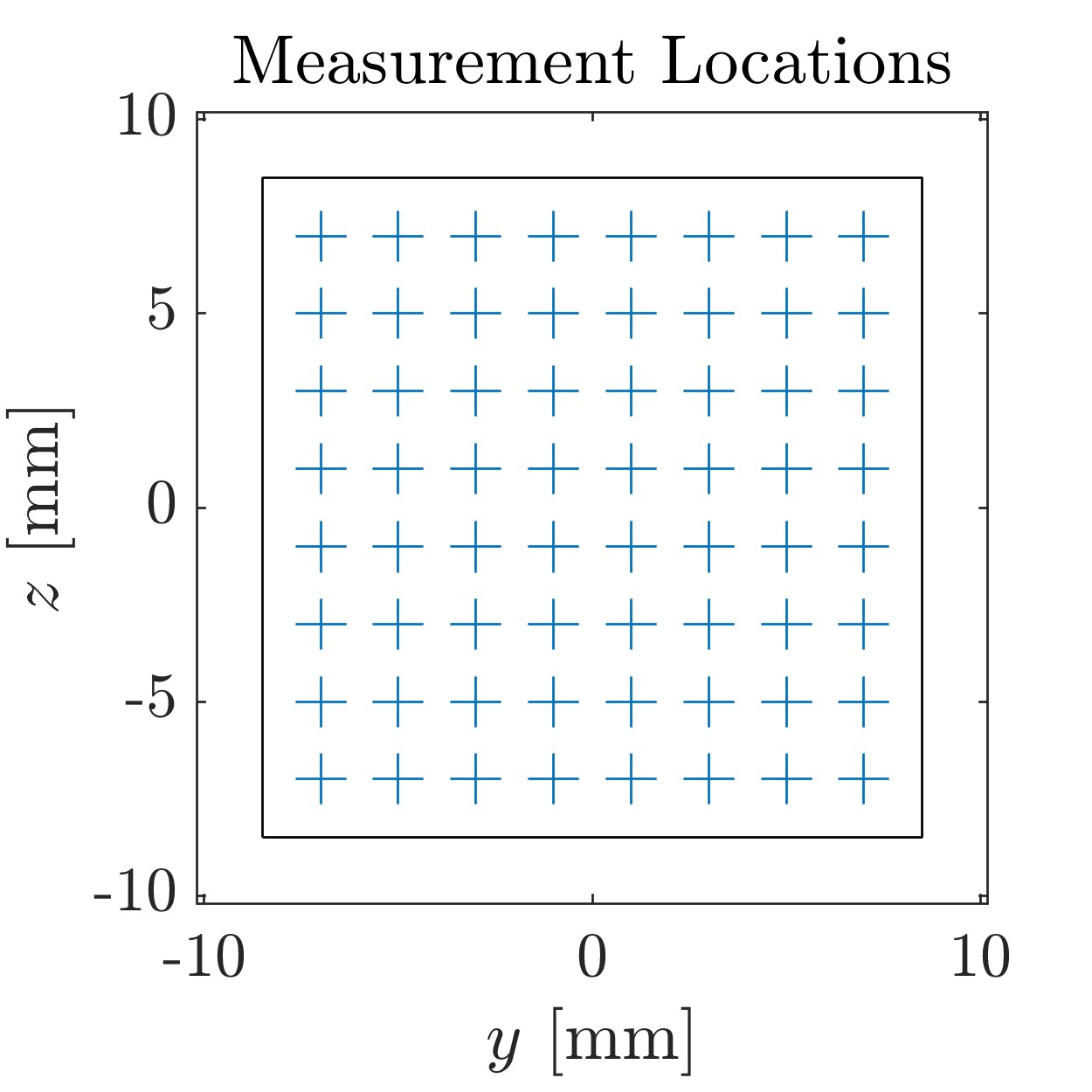}
    \put(-85,82){(c)}\\
    \includegraphics[width=0.85\linewidth]{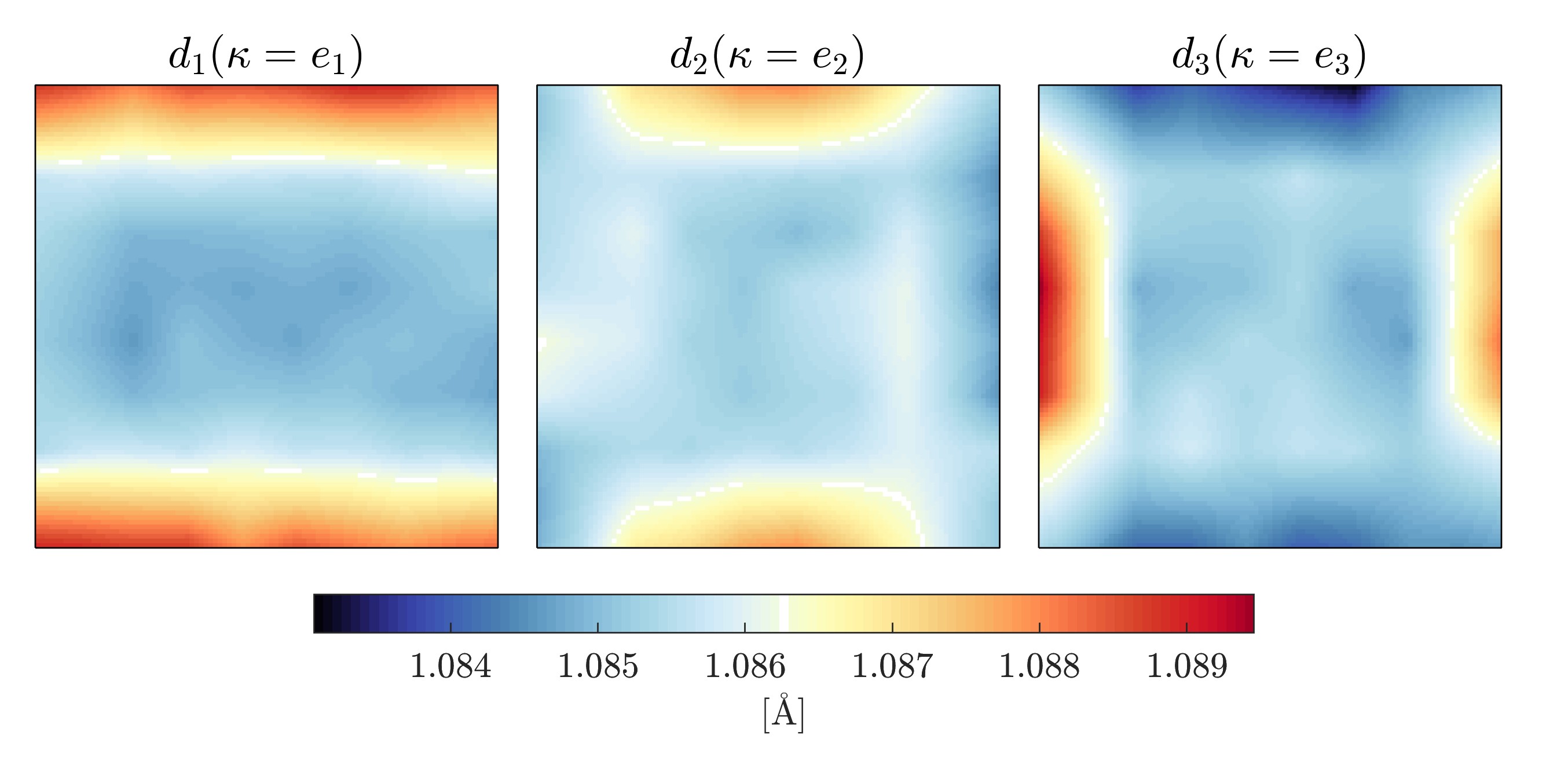}
    \put(-340,130){(d)}\\
    \includegraphics[width=0.85\linewidth]{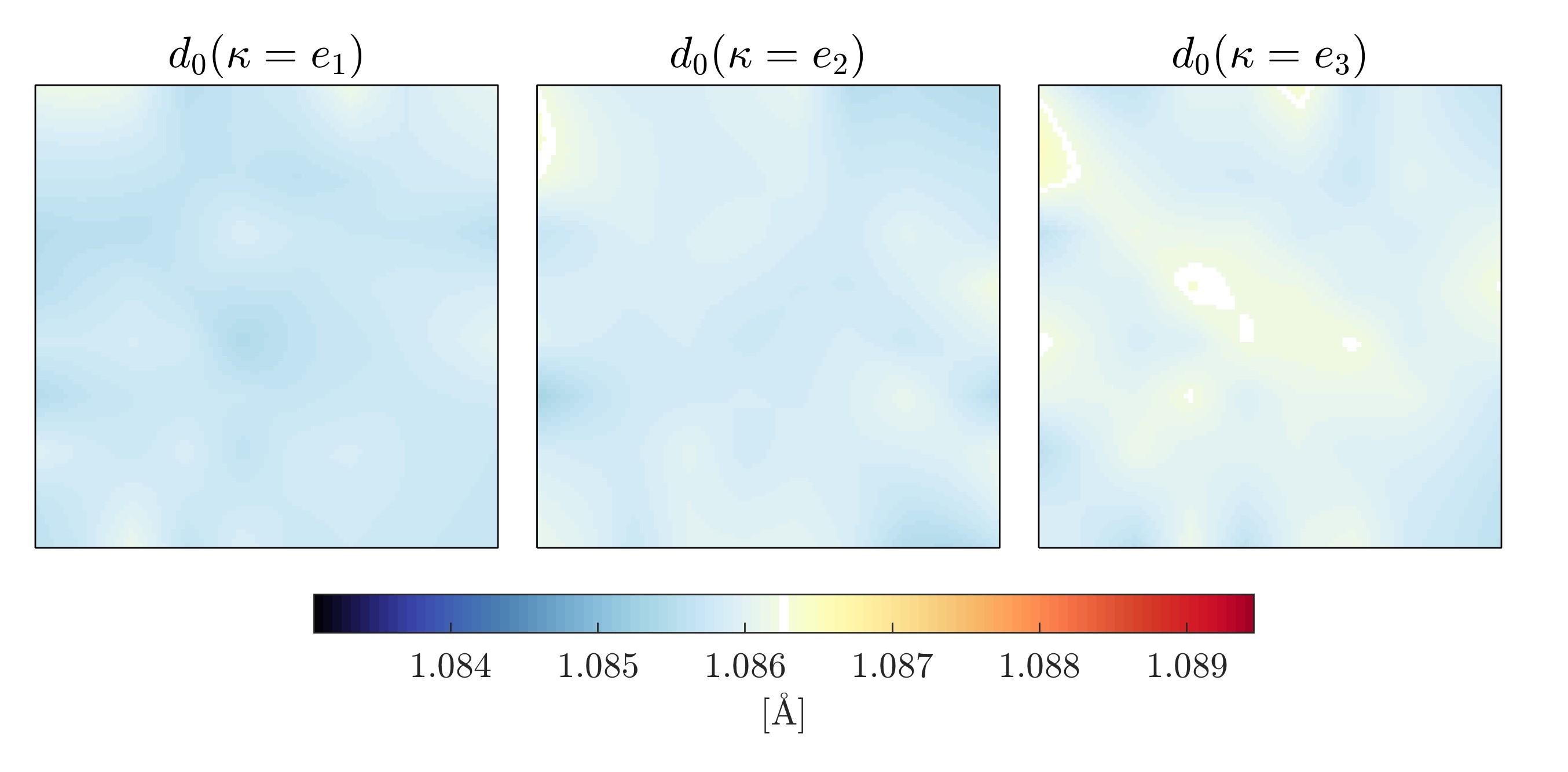}
    \put(-340,130){(e)}
    \caption{(a) A pair of nominally identical $17\times 17 \times 17$ mm Inconel cubes produced using Selective Laser Melting \cite{wensrich_residual_2024}  (b) Cube 1 was the subject of residual stress measurements over a central cross section using the KOWARI strain diffractometer at the Australian Centre for Neutron Scattering (ACNS) at the Australian Nuclear Science and Technology Organisation (ANSTO).  The schematic shows the geometry for the measurement of the $\epsilon_{11}$ component ($\kappa=Q$).  Cube 2 was dissected using Electro-Discharge Machining (EDM) to provide a stress-free `$d_0$' comb. (c) Locations of the experimental measurements over the cross section. (d) Interpolations of the measured $d$-spacings in the coordinate directions. (e) Interpolations of the direct $d_0$ measurements made in the comb sample in each of the coordinate directions.}
    \label{fig:SLMCubes}
\end{figure}

For example, consider the additively manufactured Inconel cubes produced through Selective Laser Melting (SLM) shown in Figure \ref{fig:SLMCubes}.  These cubes were part of a previous study of residual stress generated by the SLM process in which a set of neutron based strain measurements were taken over a mid-plane within one of the samples -- see \cite{wensrich_residual_2024} and \cite{wensrich_well-posedness_2025} for details.  In total, 9 separate components of strain were measured using a $1\times 1\times 1$ gauge volume over an $8\times 8$ grid of points within this plane.  In terms of the coordinate system shown, these directions were
\[
\kappa=\Bigg\{\begin{bmatrix}1\\0\\0\end{bmatrix},
\begin{bmatrix}0\\1\\0\end{bmatrix},
\begin{bmatrix}0\\0\\1\end{bmatrix},
\frac{1}{\sqrt{2}}\begin{bmatrix}1\\\pm1\\0\end{bmatrix},
\frac{1}{\sqrt{2}}\begin{bmatrix}1\\0\\\pm1\end{bmatrix},
\frac{1}{\sqrt{2}}\begin{bmatrix}0\\1\\\pm1\end{bmatrix}
\Bigg\}
\]
Measurements from the first three of these directions are shown in Figure \ref{fig:SLMCubes}d as a `\texttt{natural}' interpolation using the MATLAB `\texttt{scatteredInterpolant}' intrinsic function.

Also shown is an equivalent set of measurements over a $7\times 8$ grid in a stress-relieved `comb' that was cut using EDM from an equivalent cube printed under the same conditions.  Table \ref{tab:SLMCubeCompare} shows $d_0$ estimates based on averages in each direction along with an average over all directions.  Several things are immediately apparent from these measurements;
\begin{itemize}
    \item[--]{Within the level of uncertainty, the measured $d_0$ spacings are approximately constant in each direction.  The observed standard deviations across the 56 measurements in each direction were $1.07\times 10^{-4}\AA$, $1.03\times 10^{-4}\AA$, and $1.21\times 10^{-4}\AA$.  These are slightly higher than the measured `peak-fit' uncertainty (typically $\sim 7\times 10^{-5}$\AA), but not significantly so.}
    \item[--]{Across the different directions there is a small but significant difference between the averages.  Relatively speaking, the measured $d_0$ in the $e_3$ direction was $1.2\times 10^{-4}$ and $2.2\times 10^{-4}$ larger than that of the $e_2$ and $e_1$ directions respectively (against a relative uncertainty of $1.4\times 10^{-5}$).  While well above the level of uncertainty, this anisotropy is relatively insignificant compared to the range of variation in $d$-spacing due to the residual stress. This anisotropy may also be due to the small amount of residual stress introduced by the EDM cutting process.}
\end{itemize}

Also shown in Table \ref{tab:SLMCubeCompare} is a $d_0$ estimate via \eqref{eq:d0viaForceBal} where the integrals were calculated as Riemann sums from the interpolations shown in Figure \ref{fig:SLMCubes}.  Intrinsically this estimate assumes isotropic $d_0$ and we note that the estimate lies within the range of the separate averages over each direction. 

\begin{table} [!h]
    \centering
    \begin{tabular}{lc}
       SLM Cube $d_0$ estimates & \\
       \hline\hline
       Comb Sample (averages):  & \\
       \hspace{2ex}$e_1$ - direction  & $1.085755\pm1.4\times 10^{-5}$\AA\\
       \hspace{2ex}$e_2$ - direction  & $1.085862\pm1.4\times 10^{-5}$\AA\\
       \hspace{2ex}$e_3$ - direction  & $1.085989\pm1.4\times 10^{-5}$\AA\\
       \hspace{2ex}Overall average    & $1.085869\pm1.2\times 10^{-4}$\AA\\
       \hline
       $d_0$ from force balance: & \\
       \hspace{2ex} Equation \eqref{eq:d0viaForceBal} & $1.085808\pm0.6\times 10^{-4}$\AA \\
       \hline
       $d_0$ from boundary conditions:\\
       \hspace{2ex} Figure \ref{fig:d0FromBoundary} (average over all boundaries) & $1.085810\pm 0.6 \times 10^{-4}$\AA\\
       \hline
       $d_0$ from equilibrium: \\
       \hspace{2ex} Figure \ref{fig:CubeEigenstrainData}c (average $\pm$ range/2) & $1.085792\pm 1.1 \times 10^{-4}$\AA
    \end{tabular}
    \caption{$d_0$ estimates over the symmetry plane of the SLM Inconel cube shown in Figure \ref{fig:SLMCubes}.  Comb samples averages refer to direct measurements and these are compared to estimates from a force balance over the cross section (see Section \ref{sec:ForceBal}), an average of reconstructions over the boundary using the traction-free boundary condition (see Section \ref{sec:d0FromBC}) and the average of a full reconstruction based on equilibrium on symmetry planes (see Section \ref{sec:Equilibrium}).}
    \label{tab:SLMCubeCompare}
\end{table}

\subsection{$d_0$ from boundary conditions}\label{sec:d0FromBC}
It is often the case that some or all of the surface of our sample is free from applied load. In terms of stress, this can be expressed as the boundary condition
\[
\sigma\cdot n=0.
\]
So together with Hooke's law, any such point has three additional equations that can be included in $[K]$;
\[
[K]_{(m+3)\times7}[d_0\epsilon]_{7\times1}=[\mathcal{D}]_{(m+3)\times 1}
\]
\[
\begin{bmatrix}
\kappa_{(1)1}^2 & \kappa_{(1)2}^2 &\kappa_{(1)3}^2 & 2\kappa_{(1)1}\kappa_{(1)2} & 2\kappa_{(1)1}\kappa_{(1)3} & 
2\kappa_{(1)2}\kappa_{(1)3} &1\\
\kappa_{(2)1}^2 & \kappa_{(2)2}^2 &\kappa_{(2)3}^2 & 2\kappa_{(2)1}\kappa_{(2)2} & 2\kappa_{(2)1}\kappa_{(2)3} & 
2\kappa_{(2)2}\kappa_{(2)3} &1\\
 & & & \cdots & & &\\
 \kappa_{(m)1}^2 & \kappa_{(m)2}^2 &\kappa_{(m)3}^2 & 2\kappa_{(m)1}\kappa_{(m)2} & 2\kappa_{(m)1}\kappa_{(m)3} & 
2\kappa_{(m)2}\kappa_{(m)3} &1\\
(1-\nu)n_1 & \nu n_1 & \nu n_1 & (1-2\nu)n_2 & (1-2\nu)n_3 & 0 & 0\\
\nu n_2 & (1-\nu) n_2 & \nu n_2 & (1-2\nu)n_1   & 0 & (1-2\nu)n_3 & 0\\
\nu n_3 & \nu n_3 & (1-\nu)n_3 & 0 & (1-2\nu)n_1 & (1-2\nu)n_2 & 0\\
\end{bmatrix}
\begin{bmatrix}
d_0\epsilon_{11}\\
d_0\epsilon_{22}\\
d_0\epsilon_{33}\\
d_0\epsilon_{12}\\
d_0\epsilon_{13}\\
d_0\epsilon_{23}\\
d_0
\end{bmatrix}= \begin{bmatrix}
    d_{\kappa_{(1)}}\\
    d_{\kappa_{(2)}}\\
    \cdots\\
    d_{\kappa_{(m)}}\\
    0\\
    0\\
    0
\end{bmatrix}
\]

Generally speaking, this is usually enough to make $[K]$ full rank and allow us to solve for a unique least squares `best fit' for $d_0$ (and the corresponding strain) using the Moore-Penrose pseudo-inverse from before.  Note that every such point on a free boundary can potentially provide a separate estimate of this form.

While this intrinsically solves the `$d_0$ problem', there is some inherent difficulty in its direct application.  The problem lies in the fact that partial illumination issues dictate that the gauge volume should always be wholly contained within the sample.  Hence measurements \emph{at} the boundary are never directly available.  However, we can proceed (cautiously) with extrapolations.  

\begin{figure}
    \centering
    \includegraphics[width=0.9\linewidth]{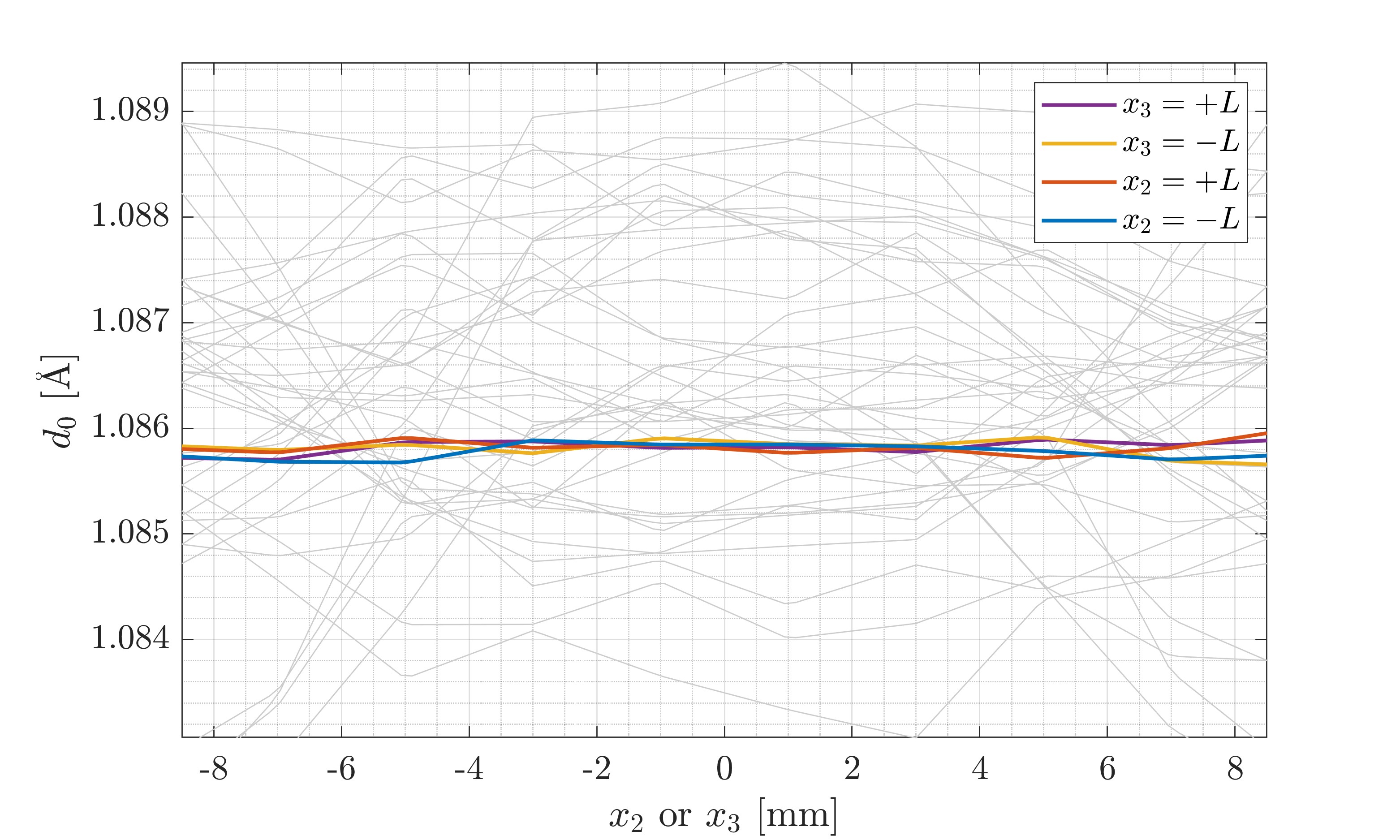}
    \caption{Calculated $d_0$ distributions over the four boundaries on the mid-plane of the SLM Inconel cube in Figure \ref{fig:SLMCubes}.  These distributions were calculated from the traction-free boundary condition based on `\texttt{natural}' extrapolations of the measured $d$-spacings using the MATLAB `\texttt{scatteredInterpolant}' intrinsic function.  Grey lines show these extrapolations at the boundary for scale.}
    \label{fig:d0FromBoundary}
\end{figure}

Figure \ref{fig:d0FromBoundary} shows the results of this process over the four separate segments of the boundary on the mid-plane of the SLM Inconel cube in Figure \ref{fig:SLMCubes}.  This calculation relied upon extrapolation of the measured $d$-spacings to the boundary using the same MATLAB `\texttt{scatteredInterpolant}' as before.  The grey lines in the Figure show this extrapolated data to give a sense of scale. Note that the scale on the vertical axis is the same as the colour scale in Figure \ref{fig:SLMCubes}d and \ref{fig:SLMCubes}e.

It is clear from Figure \ref{fig:d0FromBoundary} that the reconstructed $d_0$ is consistent across each of the four boundaries and relatively constant.  The average over all of the boundaries was $d_0=1.085810\pm 0.6 \times 10^{-4}\AA$; well within the range observed in the direct $d_0$ measurements from the comb sample.

It is also worth noting that this approach is often useful as a \emph{quasi}-direct method for measuring $d_0$ in either thin/planar samples, or specially prepared thin planar coupons in place of stress-free combs or cubes. The plane-stress nature of these types of samples allows for the calculation of $d_0$ in an entirely similar way to the above.

\section{$d_0$ from point-wise equilibrium}

The previous section detailed how boundary conditions on external surfaces and/or force balances on section planes within a sample can resolve the `$d_0$ problem' without relying on direct measurements from coupons.  In the following section we develop this idea further to explore techniques for point-wise calculation of $d_0$ based equilibrium within the sample.  

As detailed in Section \ref{Sec:Theory}, equilibrium gives us a reliable set of constraints that must be satisfied at every point within our sample.  If used carefully, these constraints can resolve the $d_0$ problem and allow for the reconstruction of $d_0$ distributions within our sample.  
In \ref{Sec:ApndxB} we show this task is mathematically possible in a general sense if sufficient data is available. Unfortunately this general version requires three-dimensionally distributed measurement locations throughout the sample which is rarely available.  However there are a number of special cases that are commonly met `in the wild' where this approach is appropriate and useful.  We present two scenarios; 
\begin{enumerate}
    \item Axisymmetric systems, and,
    \item Measurement on symmetry planes and two-dimensional systems.
\end{enumerate}

\subsection{$d_0$ from equilibrium in axisymmetric systems}

Axial symmetry refers to cylindrical geometries where all variables are a function of radius alone.  This is a common assumption for many cylindrical samples (e.g. \cite{stacey_measurement_1985,abbey_reconstruction_2012,wensrich_measurement_2012,wensrich_well-posedness_2025,jahed_axisymmetric_1997}), including all of the round-robin `ring-and-plug' reference systems that are routinely passed around by instrument scientists working on the various residual stress diffractometers around the world (e.g. \cite{daymond_analysis_2002,youtsos_vamas_2000}).

The axisymmetric assumption implies that all deformation is radial and hence the principal directions for stress and strain align with the cylindrical coordinate system ($r,\theta$ and $z$).  In this case, the governing equations from equilibrium \eqref{eq:Equilbrium} reduce to
\begin{equation}\label{eq:AxiSymEqui}
    \frac{d\sigma_{rr}}{dr}+\frac{\sigma_{rr}-\sigma_{\theta\theta}}{r}=0,
\end{equation}
and the total strain/displacement relationship becomes
\[
\varepsilon_{rr}=\frac{d\chi}{dr} \text{ and } \varepsilon_{\theta\theta}=\frac{\chi}{r}
\]
where $\chi(r)$ is the radial displacement within the sample. In the axial $z$-direction, there are two limiting cases;
\begin{enumerate}
    \item{The infinitesimally `thin' plane-stress case where $\sigma_{zz}=0$ and $\epsilon_{zz}$ is defined by $\sigma_{rr}$ and $\sigma_{\theta\theta}$ through Hooke's law, and,}
    \item{The infinitely `thick/long' plane-strain case where $\varepsilon_{zz}=0$ and $\sigma_{zz}$ is defined by $\epsilon_{zz}=-\epsilon_{zz}^*$, $\epsilon_{rr},\epsilon_{\theta\theta}$ and Hooke's law.}
\end{enumerate}
In between these two cases is the `finitely long' generalised plane strain approximation where we can assume $\varepsilon_{zz}$ is constant as long as we are sufficiently far from the ends of the sample. Boundary conditions and constraints for this system are based on continuity at the origin
\[
\chi(0)=0,
\]
force balance in the $z$-direction
\[
\int_0^R r\sigma_{zz} \text{d}r=0,
\]
and a traction-free boundary
\[
\sigma_{rr}(R)=0,
\]
where $R$ is the outer radius of the sample.

Say we have known (measured) radial distributions of $d$-spacings in the $r,\theta$ and $z$ directions.
We would like to calculate the radial distributions of the three principal stress components along with the $d_0$ distribution using these equations.  At the boundary, we can already compute $d_0$ through the zero-traction constraint.  This provides
\begin{equation}\label{eq:d0FromBdry}
d_0(R)=\frac{d_*(R)}{1+3\alpha},
\end{equation}
where $d_*=(1+\alpha)d_r+\alpha d_\theta+\alpha d_z$.

Everywhere else, using Hooke's law we can write \eqref{eq:AxiSymEqui} in terms of the known $d$-spacings as
\[
\frac{d}{dr}\Big(\frac{d_*}{d_0}\Big)+\frac{1}{r}\frac{d_r-d_\theta}{d_0}=0.
\]
Rearranging, this gives a first order ordinary differential equation for $d_0$ of the form
\[
\frac{1}{d_0}\frac{d}{dr}(d_0)=\frac{1}{d_*}\Big(\frac{d}{dr}(d_*)+\frac{d_r-d_\theta}{r}\Big),
\]
with solution
\begin{equation}\label{eq:AxiSymd0}
d_0(r)=d_0(R) \exp\Bigg[\int_R^r \frac{1}{d_*}\Big(\frac{d}{dr}(d_*)+\frac{d_r-d_\theta}{r}\Big) \text{d}r\Bigg].
\end{equation}

This explicit solution establishes both the existence and uniqueness of a $d_0$ reconstruction from this type of data. However it is rare that we would have enough data to use it directly without some form of regularisation.  None-the-less, we can proceed with numerical schemes with confidence that the underlying problem has no ambiguity.  So now the question becomes;

\vspace{2ex}
\emph{How do we best reconstruct an axisymmetric $d_0$ distribution from limited and potentially noisy data?}
\vspace{2ex}

Once again, we demonstrate approaches to this by way of example;

\subsection{Example 1: Axisymmetric ancient Roman medical probes}

\begin{figure}
    \centering
    \includegraphics[width=0.7\linewidth]{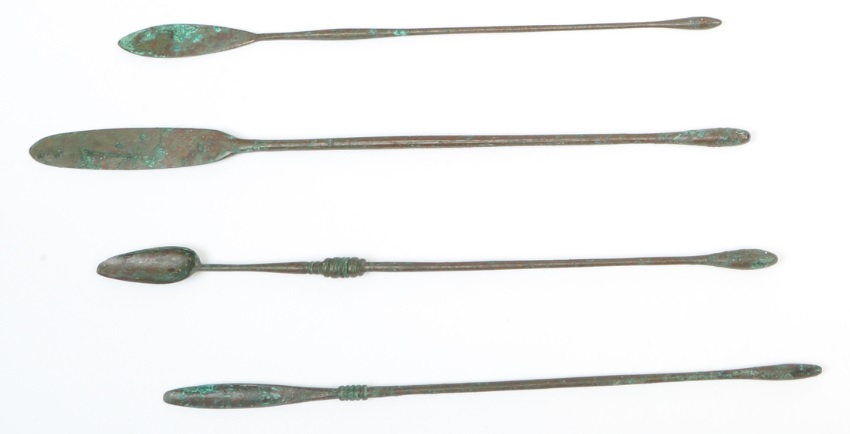}
    \caption{Four ancient Roman medical probes examined in Wensrich \emph{et al.} \cite{wensrich_well-posedness_2025}.  As a part of this study, radial distributions of $d$-spacing in the radial, hoop and axial were measured at the midpoint of the cylindrical section of each probe using the KOWARI strain diffractometer at ANSTO.  Data from the probe second from the top is shown in Figure \ref{fig:Sample3} below.}
    \label{fig:RomanProbes}
\end{figure}

Wensrich \emph{et al.} \cite{wensrich_well-posedness_2025} discusses an experimental analysis of residual stress within a set four ancient bronze medical tools using the KOWARI instrument at ANSTO (see Figure \ref{fig:RomanProbes}).  Each tool/sample features a long cylindrical section which is assumed to be axisymmetric -- at least for the sake of this analysis.

The experiment involved measuring radial distributions of lattice spacing in the radial, hoop and axial directions within each sample.  This was done using two different gauge volumes; $0.3 \times 0.3 \times 20$mm for the radial and hoop components (20mm in the axial direction), and $0.3 \times 0.3 \times 0.5$mm for the axial measurement.  The two different gauge volumes normally require two different $d_0$ values, however the relative change was determined through measurements from the same copper reference powder in each configuration.  In the data presented below, the axial $d$-spacing $d_z$ has been adjusted to account for this change.  Measurements were made as axisymmetric averages by rotating the sample during data collection.

\begin{figure}
    \centering
    \includegraphics[width=0.8\linewidth]{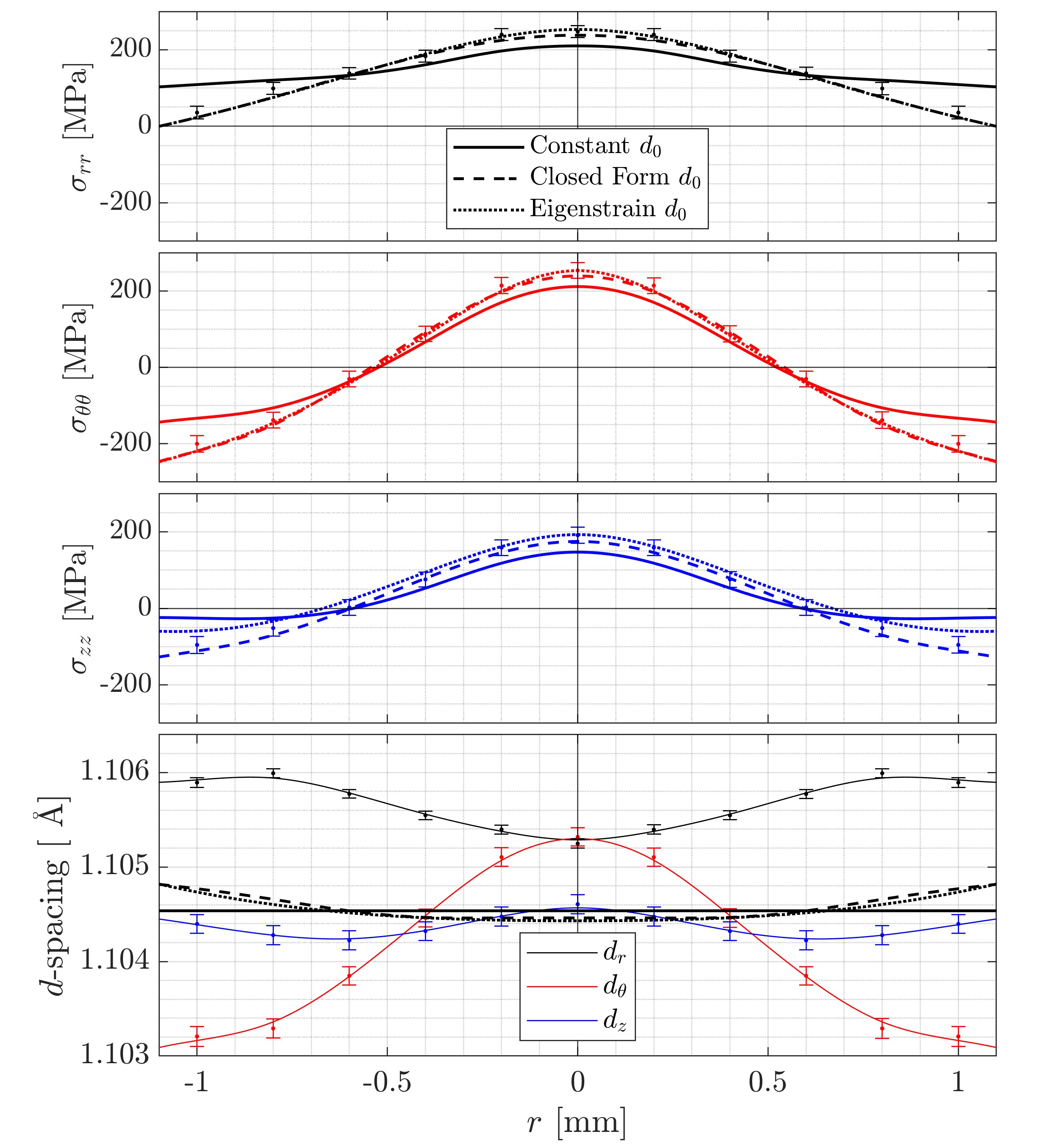}
    \put(-300,320){(a)}
    \put(-300,250){(b)}
    \put(-300,180){(c)}
    \put(-300,110){(d)}
    \caption{Residual stress predictions in the third sample shown in Figure \ref{fig:RomanProbes}.  (a),(b), and (c) Show the radial, hoop and axial components of residual stress based on an assumption of constant $d_0$ (solid lines), a direct calculation of $d_0$ from \eqref{eq:AxiSymd0} (dashed lines) and a calculation of $d_0$ based on consistency with the eigenstrain model (dotted lines).  Data markers show stress based on the eigenstrain estimate of $d_0$. (d) Shows the raw $d$-spacings measured on KOWARI along with the predicted $d_0$ distributions.}
    \label{fig:Sample3}
\end{figure}
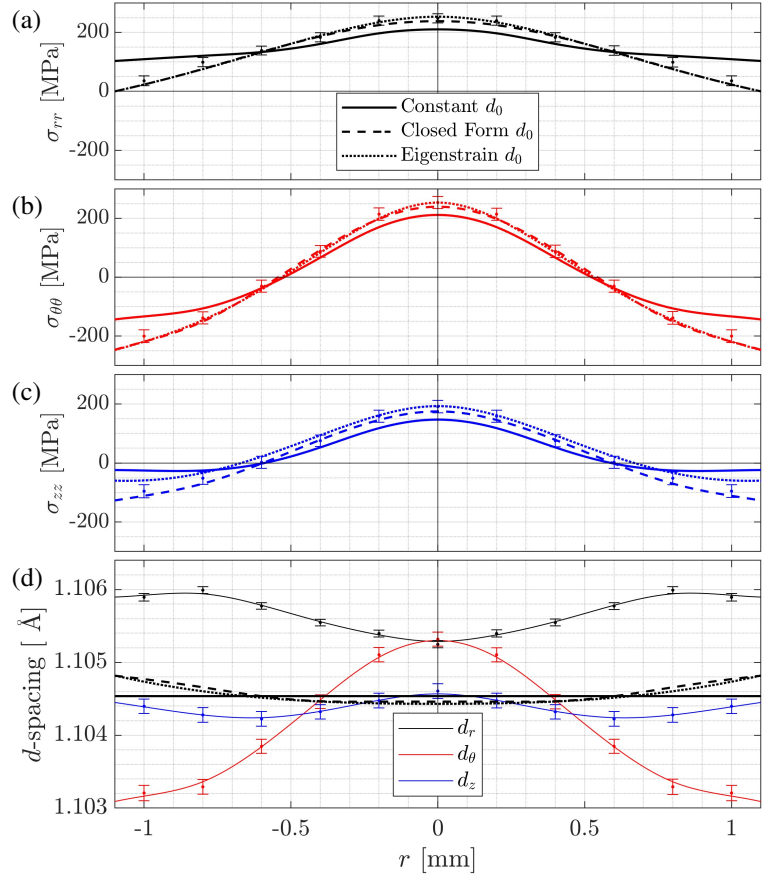

Focusing specifically on the third sample in the image ($R=1.1$mm), the measured $d$-spacings are shown as data points in Figure \ref{fig:Sample3}d (note that the axisymmetic data is mirrored over the whole cross section).  We wish to determine strain and finally residual stress from these measurements, but there is a key issue; the museum from which these samples were borrowed would not be happy if they were sent back as a pile of tiny $d_0$ fragments. Instead, we need to calculate an appropriate $d_0$ to ensure the data is consistent with equilibrium, boundary conditions and force balance.

We begin by computing a constant $d_0$ based on \eqref{eq:d0viaForceBal}.  This can be done through a relatively straightforward process involving numerical integration of intrinsic MATLAB `smoothing splines' fitted to the data.  Assuming elastic constants for 92\% Cu, 8\% Sn bronze of $E=130$GPa and $\nu=0.34$, the resulting constant reference spacing was $d_0=1.10454\pm6\times 10^{-5}\AA$ when integrating over an axial plane (normal to $z$), and $d_0=1.10456\pm6\times 10^{-5}\AA$ when integrating over a transverse plane (normal to $\theta$).  Within the level of uncertainty, these are effectively identical estimates.  Corresponding radial, hoop and axial components of stress based on this constant $d_0$ are shown in Figure \ref{fig:Sample3}a to \ref{fig:Sample3}c as smoothing splines indicated by thick solid lines.

By design, these constant $d_0$ estimates ensure the net force in either the $z$ or $\theta$ directions are zero when integrated over the cross section.  However, we notice from Figure \ref{fig:Sample3}a that the boundary condition is not satisfied; approximately 100MPa of radial stress can be seen on the surface.  We could choose a different $d_0$ to satisfy this boundary condition (based on \eqref{eq:d0FromBdry}), but it is not possible to satisfy both conditions simultaneously with a single $d_0$.  There is also no guarantee that the predicted stress satisfies \eqref{eq:AxiSymEqui}.

Building upon this, our second approach is a relatively direct application of \eqref{eq:AxiSymd0}.  This can be done as a numerical integral, however the calculation of the integrand poses two issues;
\begin{enumerate}
    \item{The integrand includes a division by $r$ that is unstable close to the origin, and,}
    \item{We need to numerically differentiate $d_*$ using limited data.}
\end{enumerate}

Note that the first issue is strictly numerical; for smooth (differentiable) $\sigma$, the axisymmetric assumption requires
\[
\sigma_{rr}(0)=\sigma_{\theta\theta}(0), \text{ and, } \frac{d\sigma_{rr}}{dr}\Bigg|_{r=0}=\frac{d\sigma_{\theta\theta}}{dr}\Bigg|_{r=0}=0.
\]
This implies that
\[
\lim_{r\rightarrow0}\frac{\sigma_{rr}-\sigma_{\theta\theta}}{r}=0,
\]
but this may not be reflected numerically when measurement uncertainty is present.  The simplest way to address this issue is to interpolate the integrand near the origin from behaviour either side.  Like before, this can be easily done using intrinsic MATLAB smoothing splines.  Similarly, we can deal with the second issue by fitting smoothing splines to $d_*$ from which derivatives can easily be calculated.

Following this process, the $d_0$ distribution shown as the thick dashed black line in Figure \ref{fig:Sample3}d was generated. Compared with the measured $d$-spacings it is relatively constant and close to the fixed $d_0$ calculated above.  Over the entire radius, the range (i.e. max–min) of the calculated $d_0$ was $3.6\times 10^{-4}$\AA; a small number, but well above the level of uncertainty.

By design, this predicted $d_0$ distribution satisfies equilibrium and the boundary condition but the net force on the cross section is unconstrained.  Integrating over the axial cross section we find a net unbalanced compressive force of 140N which implies an average unbalanced stress of 37MPa over the cross-section. 

As detailed in \cite{wensrich_well-posedness_2025}, the eigenstrain framework gives us another approach where both of these conditions can be met.  The approach is to simultaneously reconstruct an eigenstrain field together with a $d_0$ distribution that fits the observed $d$-spacings in a `least-squares' sense.  This involves an axisymmetric version of the forward calculation \eqref{eq:ForwardCalc} where the boundary value problem simplifies to the ordinary differential equation
\[
\frac{d^2\chi}{dr^2}+\frac{1}{r}\frac{d\chi}{dr}-\frac{1}{r^2}\chi=b(r),
\]
where $\chi$ is exclusively a function of $r$,
\[
    b=\frac{d \epsilon_{rr}^*}{d r}+
    \frac{\nu}{1-\nu}\Bigg(\frac{d\epsilon_{\theta\theta}^*}{d r}+\frac{d \epsilon_{zz}^*}{d r}\Bigg)+
    \frac{1-2\nu}{1-\nu}\frac{\epsilon_{rr}^*-\epsilon_{\theta\theta}^*}{r},
\]
and the boundary conditions become
\[
\begin{split}
    \chi(0)&=0, \text{ and}\\
    \label{ZeroTraction}
    \Big[(1-\nu)\frac{d \chi}{d r}+\nu\frac{\chi}{r}\Big]_{r=R}&=\Big[(1-\nu)\epsilon_{rr}^*+\nu\epsilon_{\theta\theta}^*+\nu(\epsilon_{zz}^*-\bar{\epsilon}_{zz})\Big]_{r=R}.
\end{split}
\]
Wensrich \emph{et al.} \cite{wensrich_well-posedness_2025} gives an explicit solution to this ordinary boundary value problem in the case where $\epsilon_{rr}^*,\epsilon_{\theta\theta}^*$ and $\epsilon_{zz}^*$ are defined by arbitrary polynomials.  Armed with this solution, fitting the observed data becomes an optimisation problem where we seek to minimise sum-squared error between the measured $d$-spacings and predictions of the form
\[
\begin{split}
    d_r&=\big(1+\epsilon_{rr}(r)\big)d_0(r)\\
    d_\theta&=\big(1+\epsilon_{\theta\theta}(r)\big)d_0(r)\\
    d_z&=\big(1+\epsilon_{zz}(r)\big)d_0(r),
\end{split}
\]
by choosing polynomial coefficients for $\epsilon_{rr}^*,\epsilon_{\theta\theta}^*, \epsilon_{zz}^*$ and $d_0$.  In this endeavour, we can minimise the search space by exploiting the fact that many eigenstrain fields do not generate stress.  As per Wensrich \emph{et al.} \cite{wensrich_well-posedness_2025}, we can ignore these fields by restricting 
\[
\epsilon_{\theta\theta}^*=\frac{d}{dr}(r\epsilon_{rr}^*).
\]

Using 4$^{th}$ order polynomials for eigenstrain, Figure \ref{fig:Sample3} shows the results of this least-squares model as dotted lines.  Data markers in Figure \ref{fig:Sample3} show stress calculated at the measurement locations based on $d_0$ predicted by this method.  Note the boundary condition $\sigma_{rr}(R)=0$ is perfectly satisfied, and integration over the cross section gives a net-unbalanced axial force of 0.3N.  The calculated $d_0$ distribution approximately follows the direct calculation, but with slightly more variation; a range of $3.9\times 10^{-4}$\AA.  Once again, this is small, but above the level of uncertainty.

Comparing the three estimated stress distributions we see that they all follow the same trend; almost hydrostatic tension at the centre with compressive hoop-stress at the surface.  This is a residual stress distribution that is consistent with casting/quenching.  Looking closer, in the radial and hoop directions the direct and eigenstrain approach are very similar with the constant $d_0$ predicting less variation in stress (noting, once again, non-compliance with the boundary condition in this case).  We see the most difference between the direct and eigenstrain approaches near the outer diameter.  In this region, the eigenstrain approach predicts an axial stress between the constant-$d_0$ and direct approaches.  Given it enforces both force-balance and the boundary condition, this is to be expected.

This analysis suggests that variation in $d_0$ is important to provide consistency in the data.  Is this proof that $d_0$ varies in this sample?  Probably not, and perhaps we will never know for sure.  However we have shown that it may be possible to shed light in this area through enforcing the equilibrium constraint.

\subsection{$d_0$ from equilibrium on symmetry planes and two-dimensional systems}\label{sec:Equilibrium}

As a slightly more general application of this approach, we now consider another common scenario of two-dimensional measurements made over symmetry planes and/or planar systems (e.g. plane-stress or plane-strain).  By way of example, we will focus on two systems; 1. The additively manufactured Inconel cube featured above, and 2. A bronze dagger attributed to the ancient Persian period (circa 1000BC).  By considering equilibrium at all points within these systems we can reconstruct the distribution of $d_0$ over the plane of measurement.

Before we do this, it is important to point out that it is mathematically possible to achieve such a thing. \ref{Sec:ApndxA} provides a proof that it is possible to uniquely determine $d_0$ over a symmetry plane through a combination of equilibrium and knowledge of at least 3 orthogonal $d$-spacings on the plane -- at least in a linearised sense.  As mentioned earlier, \ref{Sec:ApndxB} goes on to prove that it is generally possible to reconstruct $d_0$ everywhere within a sample from four measured $d$-spacings, although this is perhaps less practical in terms of the extent of information required.

Both of these proofs rely on the inversion of a differential system of equations in the Fourier domain, and in principle we could attempt to carry this out directly in a numerical sense.  However, the sparse and relatively noisy nature of neutron measurements does not lend themselves well to this approach.  In this setting, the eigenstrain framework provides a natural regularisation that allows us to proceed.

We begin by building an eigenstrain model for the system in question.  With reference to \cite{wensrich_well-posedness_2025}, this can be done as follows;

First define a suitable basis for eigenstrain within the sample; $\{\varphi_i,i\in[1,n_b]\}$.  Each $\varphi_i$ is a well-defined (and differentiable) distribution of the eigenstrain tensor over the whole sample (e.g. polynomial terms, or radial bump-functions).  Using this basis we can construct a general eigenstrain through a linear combination; i.e. for a given set of coefficients $\beta_i$
\begin{equation}\label{eq:eigenstrainbasis}
\epsilon^*=\sum_{i=1}^{n_b} \beta_i\varphi_i.
\end{equation}
We note that this finite representation cannot span all possible eigenstrains; the aim is to use an efficient set of basis eigenstrains that span `enough' for our needs.  

We then `forward map' each basis eigenstrain to a corresponding stress at each of the measurement locations over the plane.  This can be done using a finite element approach to solve \eqref{eq:BVP} with the right hand sides defined by each basis eigenstrain in turn.  Collecting each forward-mapped stress (as an appropriate one-dimensional array) as the columns of a matrix $A$, we can form the system of equations
\[
[A]_{6n_p \times n_b}[\beta]_{n_b \times 1}=[\mathcal{S}]_{6n_p \times 1},
\]
where $n_p$ is the number of measurement points on the plane and $\mathcal{S}$ is a one dimensional array containing the six components of a hypothetical stress at all of these measurement locations (arranged in the same form as the columns of $A$).

As before, we can use the Moore-Penrose pseudo-inverse of $A$ to find the `least-squares' best fit for the coefficients
\[
[\beta]=[A^+][\mathcal{S}],
\]
where $[A^+]=([A]^T[A])^{-1}[A]^T$.  If necessary, we can also introduce some Tikhonov regularisation by constructing the pseudo-inverse as
\[
[A^+]=([A]^T[A]+\gamma^2[W]^2)^{-1}[A]^T,
\]
where $[W]$ is a diagonal matrix of weights or penalties associated with each basis field and $\gamma$ is a regularisation (Tikhonov) parameter. Through this approach we can penalise individual basis fields such as the high-order or high-frequency terms, whatever the case may be.

Regardless of which form is used, the coefficients $[\beta]$ correspond to a set of `best-fit' stresses at the measurement points
\begin{equation}\label{eq:EigenStrainModel}
[\mathcal{S}^+]=[A][\beta]=[A][A^+][\mathcal{S}],
\end{equation}
which is the closest version to the hypothetical stress that can be represented by our eigenstrain basis.  Most importantly, this is the closest version we can achieve with the given basis that inherently satisfies equilibrium and the boundary condition.

Armed with this model, we can estimate $d_0$ over the measurement locations by the following process;

Assuming the measurements are distributed over the $x_1$-$x_2$ plane, we start with a parameterisation of the $d_0$ distribution (linear or otherwise) of the form
\begin{equation}\label{eq:d0_parameterisation}
d_0=d_0(x_1,x_2;\{a_i,i\in(1,N)\}).
\end{equation}
We seek to find the set of coefficients $\{a_i,i\in(1,N)\}$ that best describe the observed data.

Relative to the measured $d$-spacings, any given choice of coefficients defines a set of elastic strain `measurements' at each measurement location of the form
\[
\Big\{ \epsilon_{\kappa (i)}=\frac{d_{\kappa (i)}-d_0}{d_0}, i\in(1,m) \Big\}
\]
where $d_0$ is evaluated using \eqref{eq:d0_parameterisation} at the point in question.  We can then compute the full strain tensor at each measurement location by successive application of \eqref{eq:LeastSquaresStrain}.  Through Hooke's law, these then correspond to a set of hypothetical stress tensors at each measurement location $[\mathcal{S}]$.

We can then use \eqref{eq:EigenStrainModel} to map this hypothetical stress to a `best-fit' eigenstrain model that satisfies equilibrium and the boundary conditions, which in-turn defines elastic strain at the measurement points and corresponding predictions for the measured $d$-spacings through repeated application of \eqref{eq:fitd0}.

Parameters for $d_0$ can then be found through a (non-linear) least-squares optimisation process aimed at minimising the residuals between the observed $d$-spacings and these predictions.

To illustrate this process, we present two examples as follows;

\subsection{Example 2: $d_0$ on a symmetry plane in the SLM Inconel cube}

Turning our attention back to the SLM inconel cube from Figure \ref{fig:SLMCubes}, we can determine $d_0$ over the symmetry plane as a natural extension to the eigenstrain model detailed by Wensrich \emph{et al} \cite{wensrich_well-posedness_2025}.  Briefly, this model employed a symmetric even-ordered polynomial basis for eigenstrain, but specifically excluded any nil-potent or null eigenstrain.  This allows for a large span of the eigenstrain basis while simultaneously minimising the number of basis elements. Symmetric polynomials up to a total homogeneous order of 18 were used, with details on their construction given in \cite{wensrich_well-posedness_2025}.  

Each of these basis functions was mapped to stress using a finite element model for the whole cube implemented with the MATLAB PDE toolbox.  The columns of $A$ were computed as interpolations from the finite element solutions at the measurement locations.

\begin{figure}
    \centering
    \includegraphics[width=0.49\linewidth]{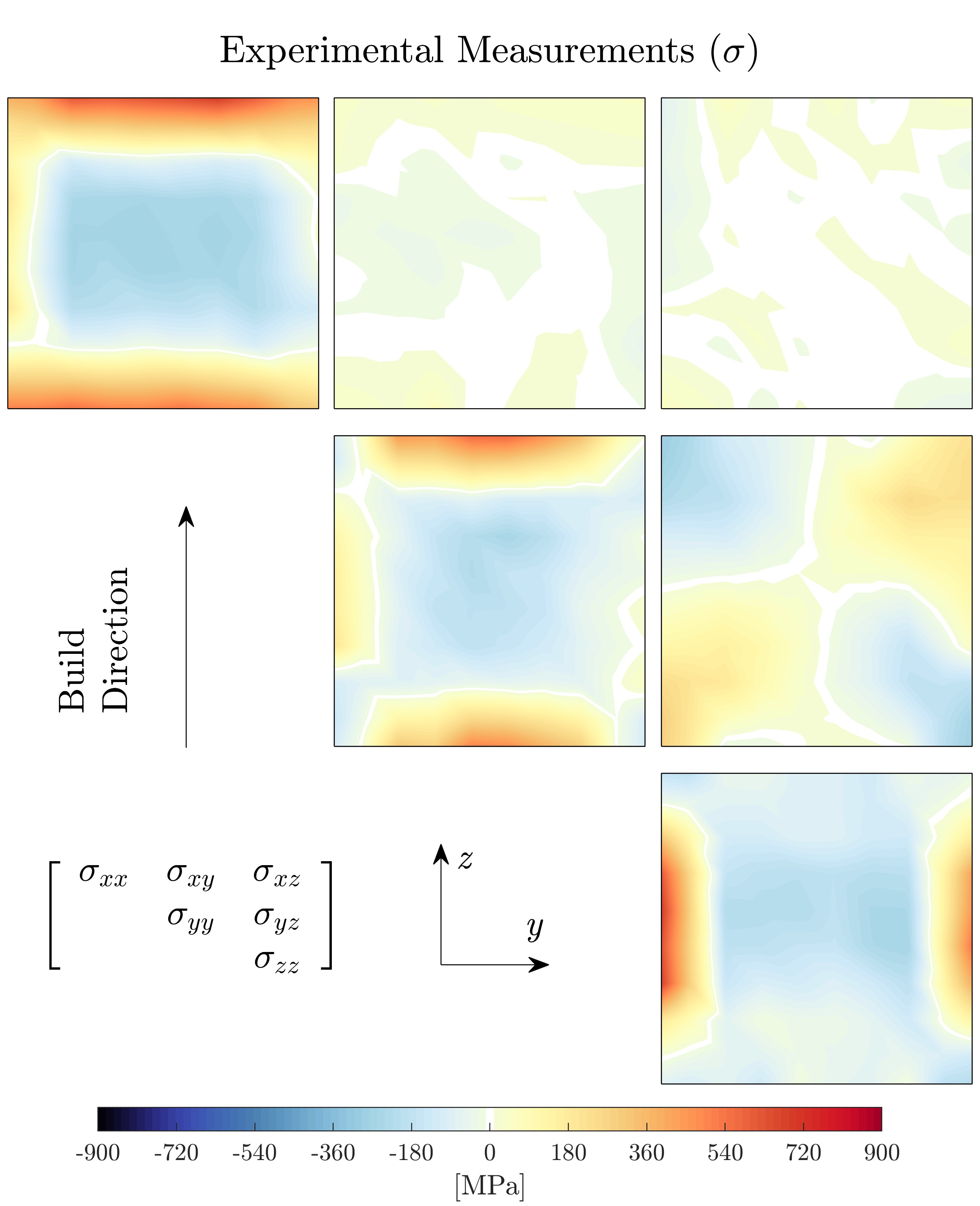}
    \includegraphics[width=0.49\linewidth]{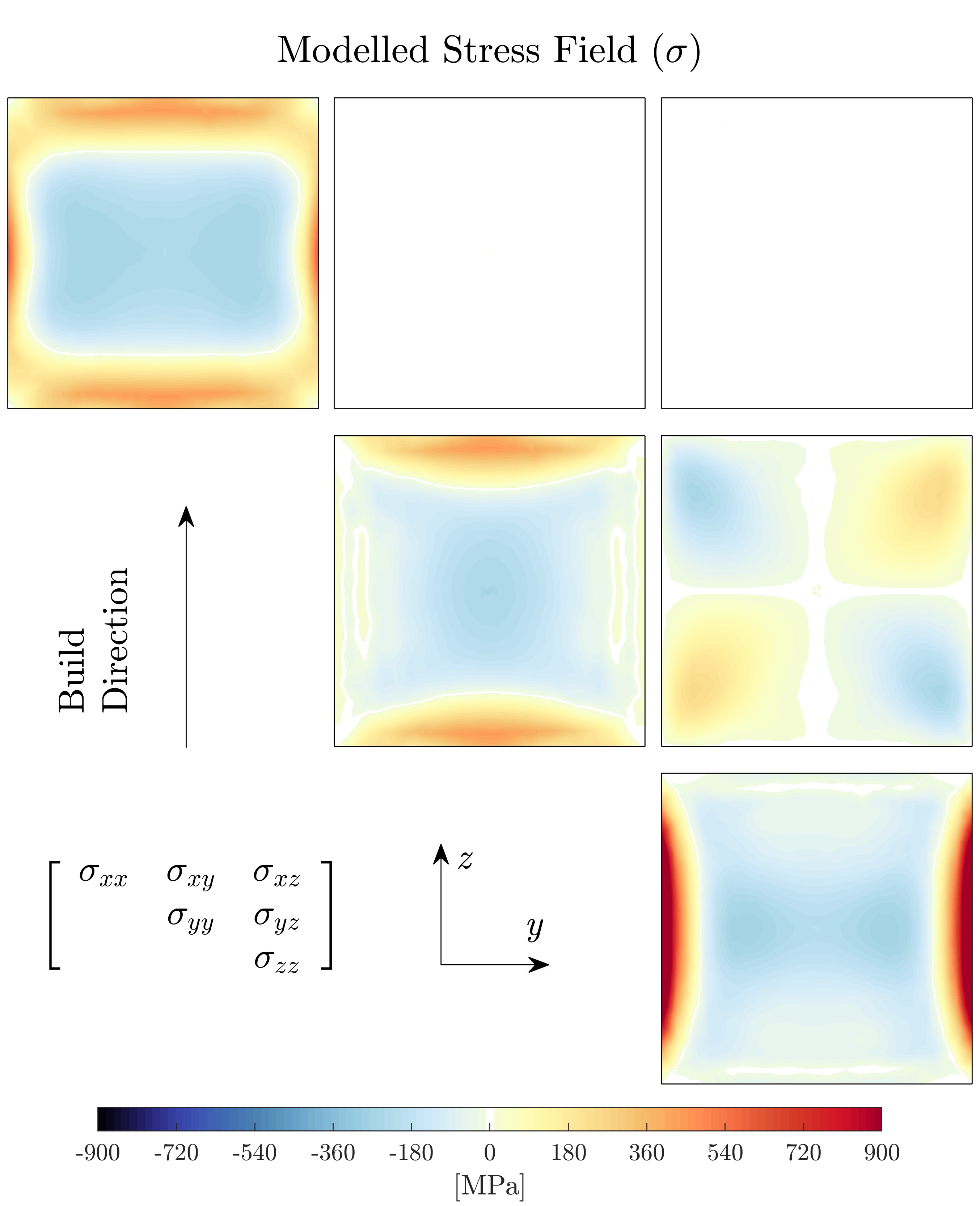}
    \put(-370,220){(a)}
    \put(-177,220){(b)}\\
    \includegraphics[width=0.37\linewidth]{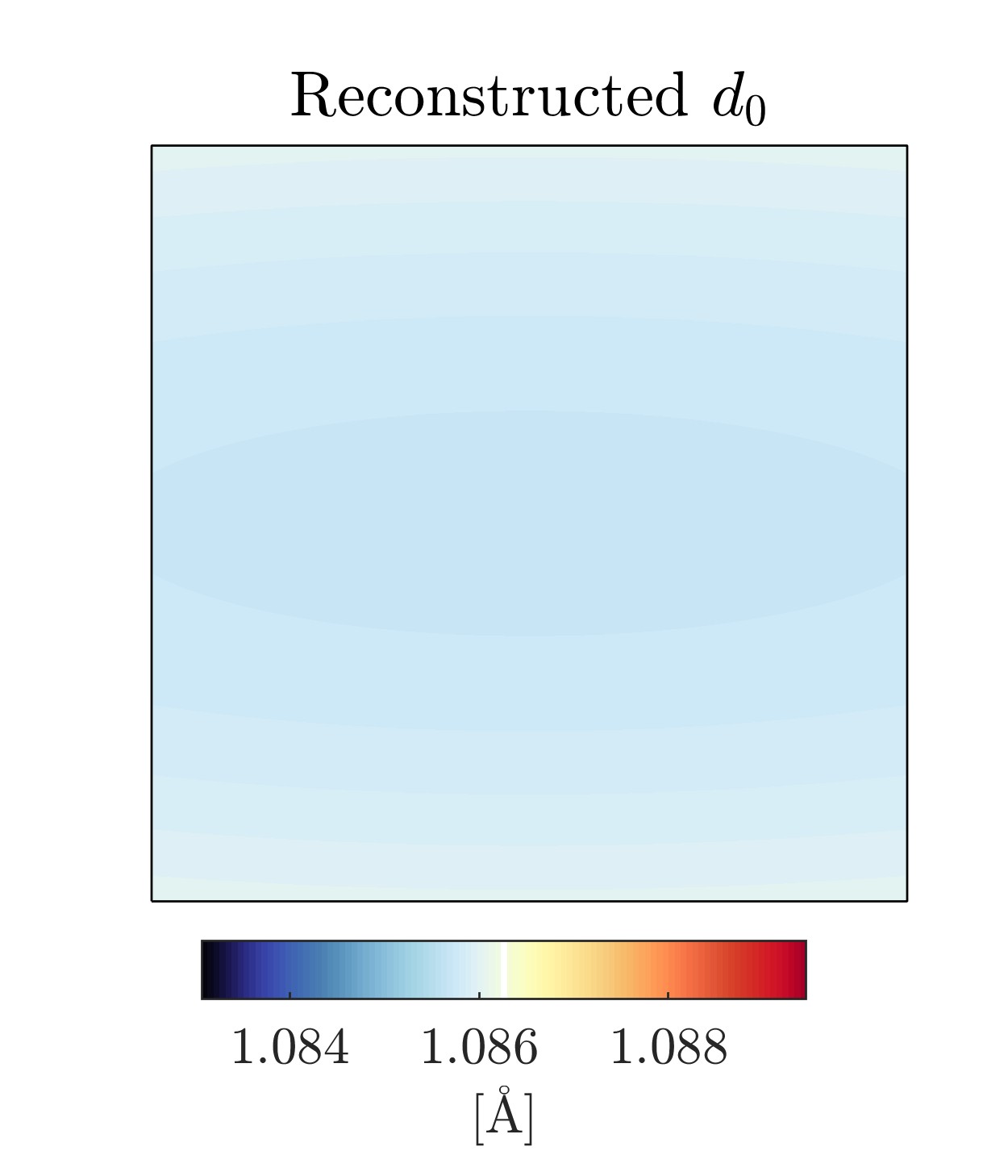}
    \put(-140,140){(c)}
    \caption{Results of the eigenstrain-based technique for reconstruction of $d_0$ over the symmetry plane in the SLM Inconel cube shown in Figure \ref{fig:SLMCubes}. (a) Interpolated components of stress based on the measured $d$-spacings. (b) The fitted eigenstrain model, and (c) The reconstructed distribution of $d_0$ over the symmetry plane shown in the same colour-scale as Figure \ref{fig:SLMCubes}d and \ref{fig:SLMCubes}e.}
    \label{fig:CubeEigenstrainData}
\end{figure}

With reference to the coordinate system in Figure \ref{fig:SLMCubes}b, and exploiting the symmetry, we can parameterise $d_0$ as a two-dimensional polynomial of even-ordered terms
\begin{equation}\label{eq:d0_representation}
d_0(x_2,x_3)=\sum_{i,j=0}^N a_{ij}x_2^{2i}x_3^{2j},
\end{equation}
for some limited value of $N$.  On this basis, we seek the coefficients $a_{ij}$ for $i,j\in[0,N]$ that minimise the discrepancy between the fitted eigenstrain model and the measured $d$-spacings. This was achieved using the Levenberg-Marquardt algorithm within the `\texttt{lsqnonlin}' function from the MATLAB Optimization Toolbox with the results of the process for $N=2$ shown in Figure \ref{fig:CubeEigenstrainData}.

Figure \ref{fig:CubeEigenstrainData} shows two different versions of the stress distribution in the cube shown as components over the symmetry plane; Figure \ref{fig:CubeEigenstrainData}a shows an interpolation over the measurement locations calculated directly from the measured $d$-spacings and fitted $d_0$, while Figure \ref{fig:CubeEigenstrainData}b shows the least-squares best fit from the eigenstrain model.  The computed distribution of $d_0$ is shown in Figure \ref{fig:CubeEigenstrainData}c with the same colour scale as the measured $d$-spacings in Figure \ref{fig:SLMCubes}.  With this model for $d_0$, the standard deviation of the residuals between predicted and measured $d$-spacings was $1.6\times 10^{-4}$; larger than the typical peak-fit uncertainty of $0.9\times 10^{-4}$, but still a reasonably good fit to the data.

Over all of the measurement locations the average $d_0$ predicted by the model was $1.085792 \AA$ with a maximum variation (range) over these same points $2.1\times 10^{-4} \AA$.  As we would expect, these values are well within the relatively constant direct $d_0$ measurements observed in the comb sample (see Table \ref{tab:SLMCubeCompare}).

\subsection{Example 3: $d_0$ in a plane-strain system}

\begin{figure}
    \centering
    \includegraphics[width=0.33\linewidth]{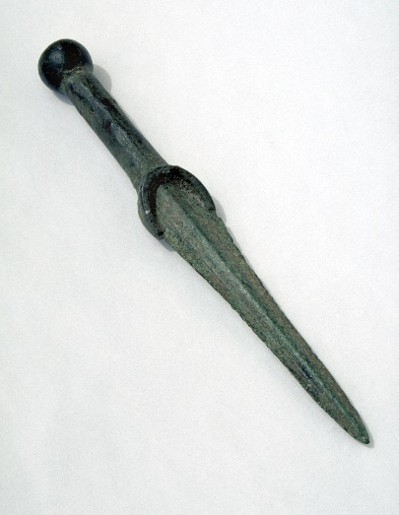}
    \includegraphics[width=0.66\linewidth]{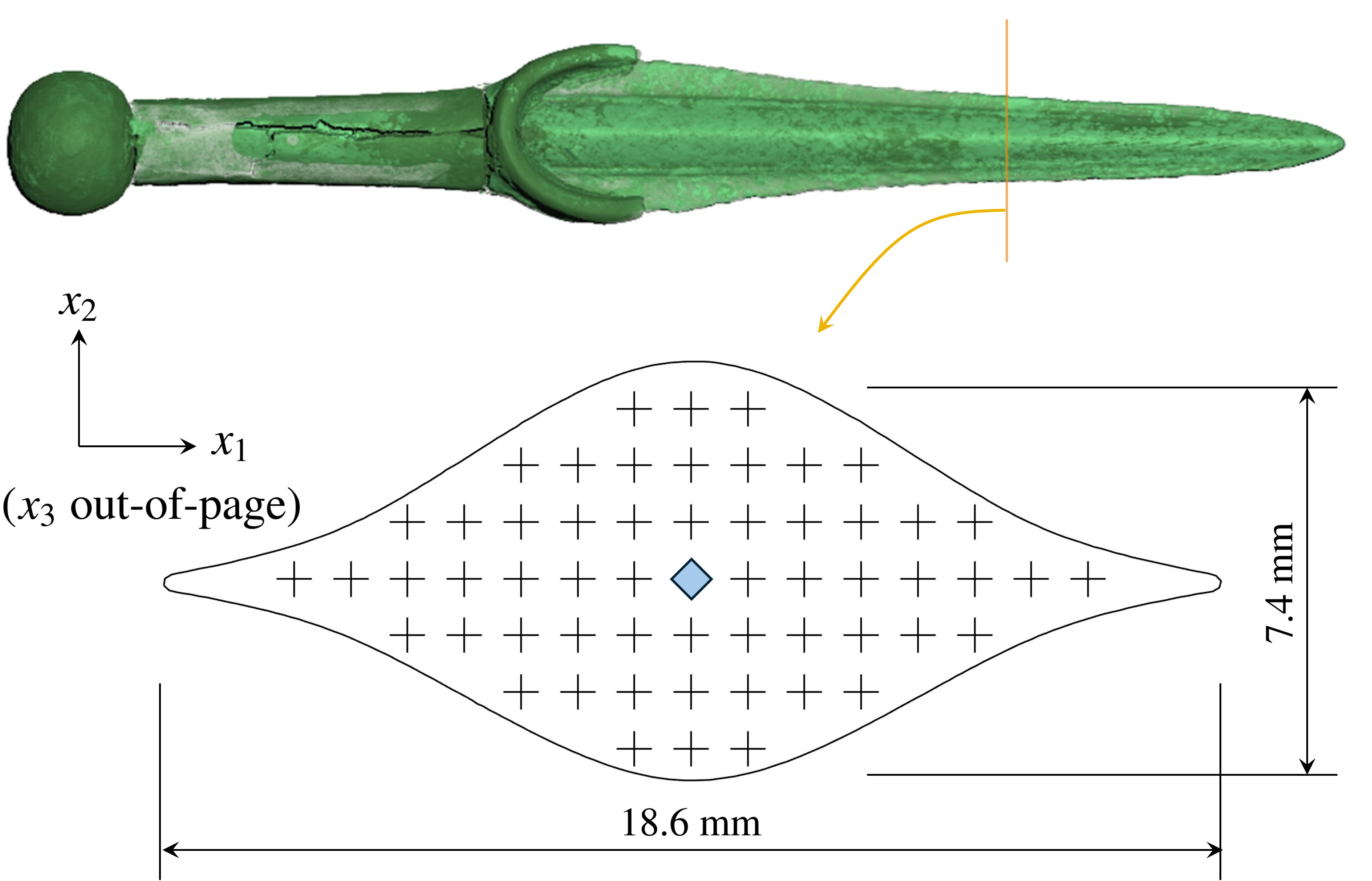}
    \caption{A bronze dagger attributed to the ancient Persian period (circa 1000BC) from the RD Milns Antiquities Museum at the University of Queensland.  This dagger was examined with the aid of the DINGO neutron imaging instrument and KOWARI strain diffractometer at the Australian Centre for Neutron Scattering (ACNS).  The upper right shows a tomographic scan from DINGO from which an indicative cross section was extracted at the measurement location (60 mm down from the tip). Over this cross section a set of $d$-spacings (axial, normal and transverse) were measured using KOWARI.  Measurement locations are indicated along with a gauge volume at the centre for scale.}
    \label{fig:PersianDagger}
\end{figure}

As a further example of this approach, we now consider a set of diffraction measurements acquired from a bronze dagger attributed to the ancient Persian period (circa 1000BC) shown in Figure \ref{fig:PersianDagger} whose authenticity remains the subject of ongoing investigation.  In collaboration with the RD Milns Antiquities Museum at the University of Queensland, this dagger underwent examination at the Australian Centre for Neutron Scattering in the form of combined tomographic imaging on the DINGO neutron imaging instrument and diffraction-based strain scanning on the residual stress diffratometer KOWARI.  The aim of this work was to gain insight into the materials and manufacturing techniques used to produce this dagger.  Although for the context of this paper, we are primarily concerned with analysis techniques for estimating residual stress.

As with the Roman medical tools, destructive techniques to establish $d_0$ are completely inappropriate and we need to resort to computational techniques.  Bronze is also a material that is notorious for $d_0$ variation (due to dislocation density and variation in composition), and so we will apply the technique discussed earlier to establish this distribution from the equilibrium constraint.

\begin{figure}
    \centering
    \includegraphics[width=0.45\linewidth]{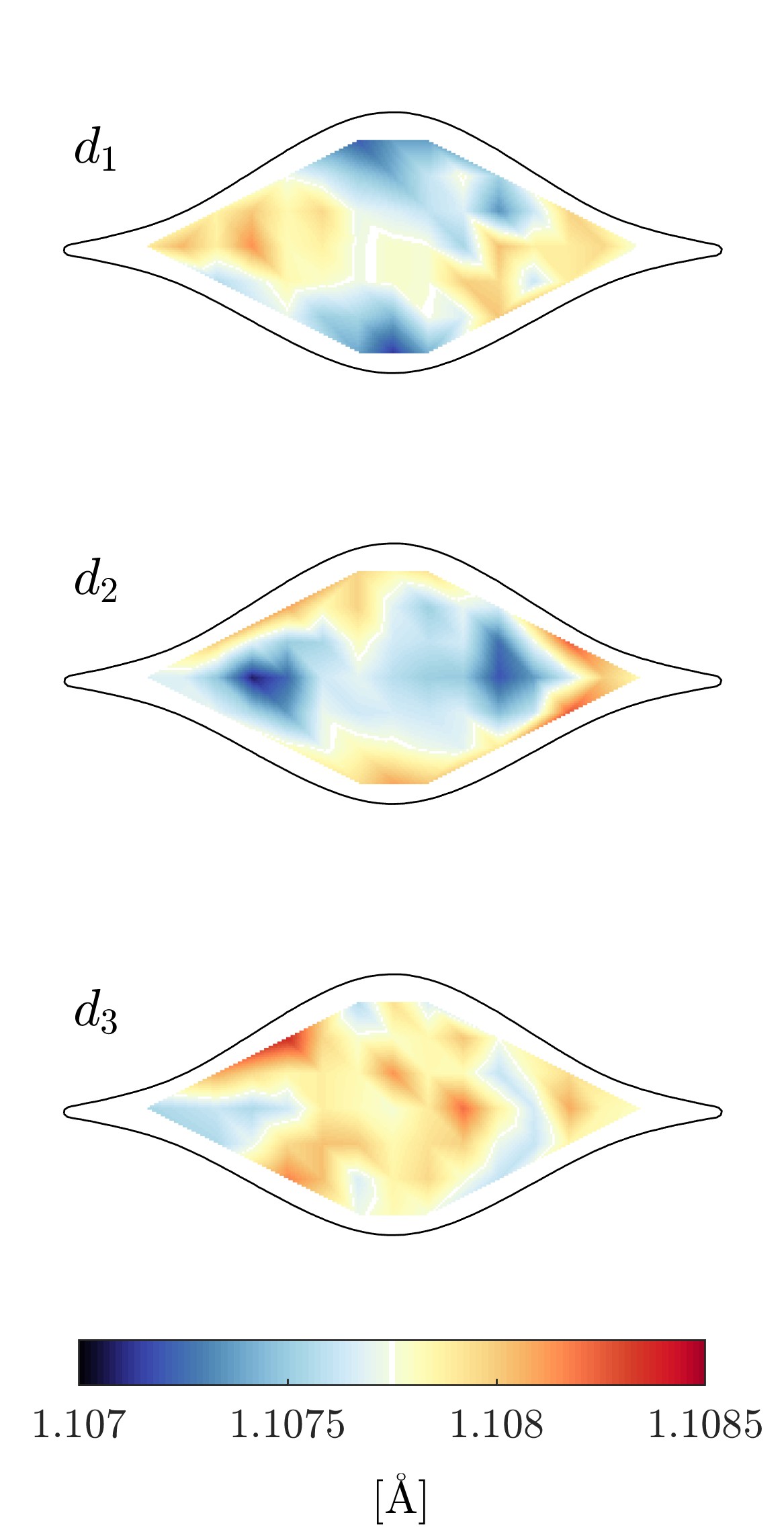}
    \includegraphics[width=0.45\linewidth]{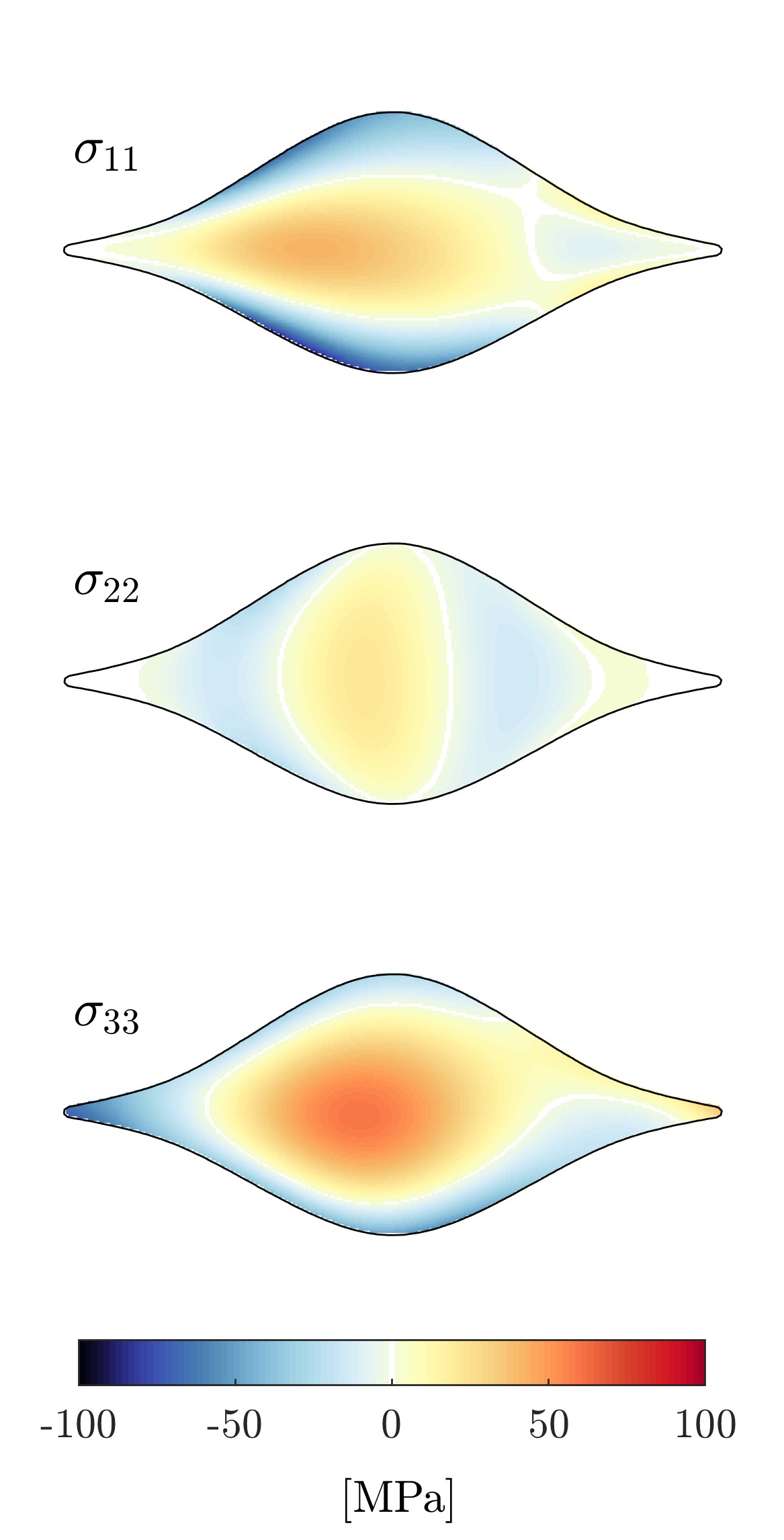}
    \put(-340,320){(a)}
    \put(-165,320){(b)}\\
    \includegraphics[width=0.45\linewidth]{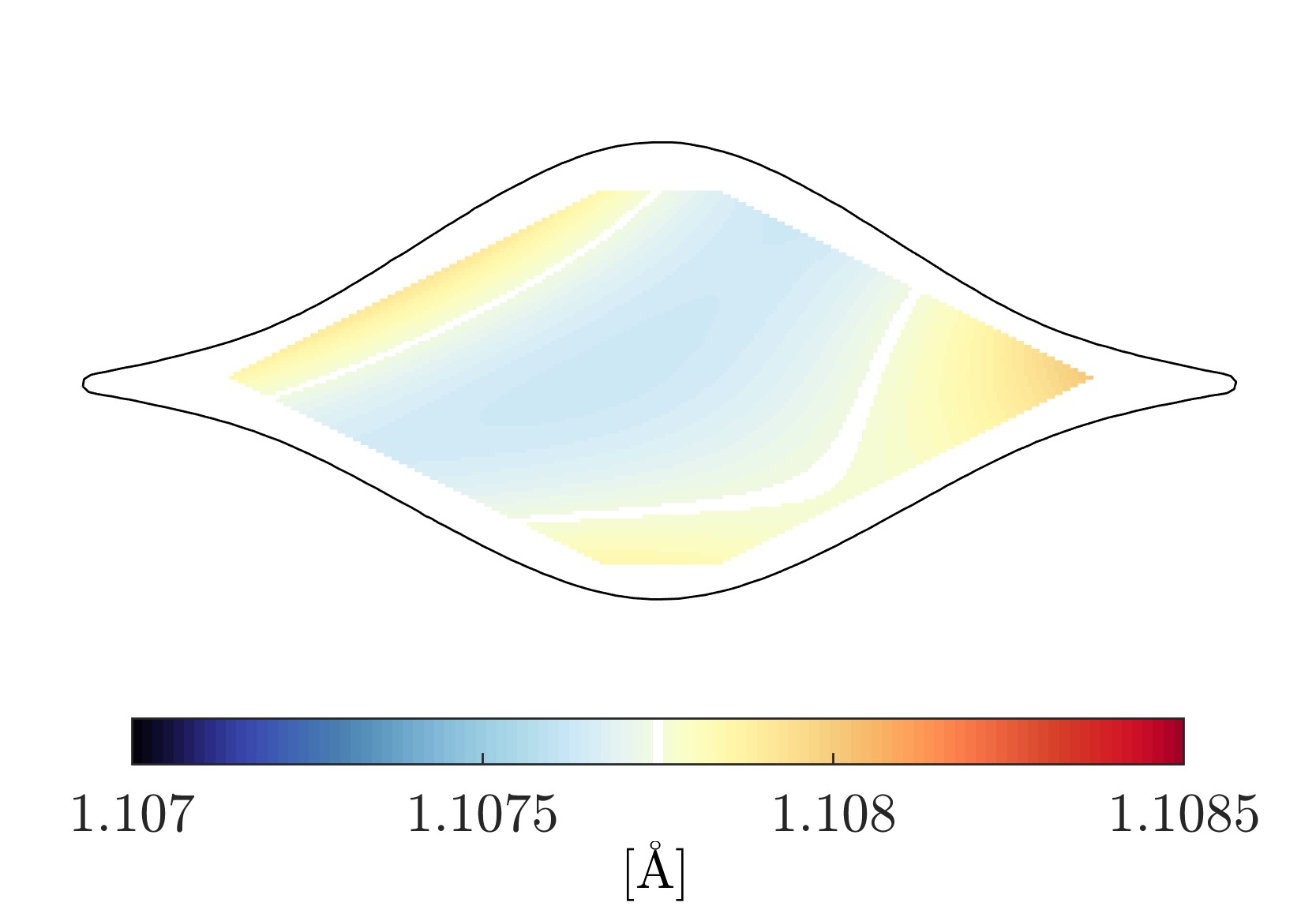}
    \includegraphics[width=0.5\linewidth]{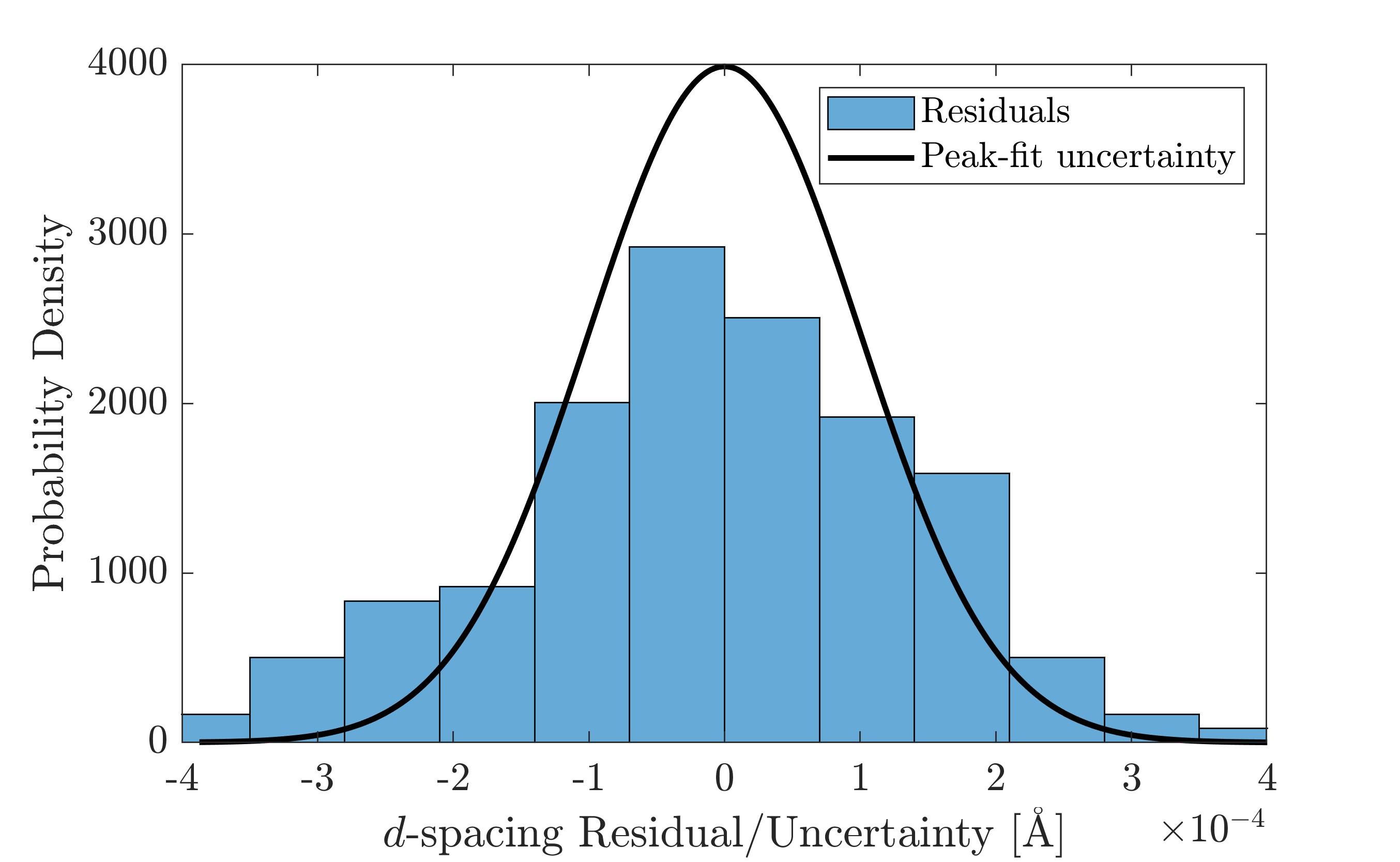}
    \put(-340,120){(c)}
    \put(-165,120){(d)}
    \caption{Results from the KOWARI strain scanning experiment on the Persian dagger.  (a) Measured $d$-spacings in the transverse $d_1$, normal $d_2$, and axial $d_3$ directions.  The colourmaps shown are interpolations from the data points shown in Figure \ref{fig:PersianDagger}.  (b) The best-fit eigenstrain solution for residual stress in the dagger. (c) The corresponding calculated $d_0$ over the cross section, and (d) Residuals between the measured $d$-spacings and predicted values from the best-fit eigenstrain model shown alongside the typical peak-fit uncertainty.}
    \label{fig:DaggerData}
\end{figure}

Using the KOWARI diffractometer, $d$-spacings were measured over a set of 57 points on a planar cross section 60mm from the tip of the dagger.  Figure \ref{fig:PersianDagger}c shows the locations of the points relative to an approximate boundary from the tomographic scan at this location in the blade. At each of these points the $d$-spacing was measured in the transverse $(d_1)$, normal $(d_2)$ and axial $(d_3)$ directions.  A gauge volume of $0.5\times 0.5 \times 12$mm was used for the normal and transverse measurements and $0.5\times 0.5 \times 0.7$mm was used in the axial direction.  The size of the gauge volume is shown to scale in Figure \ref{fig:PersianDagger}c.  A $d$-spacing offset/adjustment between these two gauge volumes was determined through the use of a copper reference powder. Count times were adjusted to maintain a target peak-fit uncertainty; this was typically $0.8\times10^{-4}\AA$ in the normal and transverse directions and $1\times10^{-4}\AA$ in the axial direction.  The resulting measured $d$-spacing distributions are shown in Figure \ref{fig:DaggerData}a as interpolations restricted to within the measurement grid.  

Through numerical integration of these interpolated distributions (noting that the axial direction is normal to the plane), \eqref{eq:d0viaForceBal} predicted a constant value of $d_0=1.10778\AA$.  We can compare this to a computed distribution calculated using the procedure in Section \ref{sec:Equilibrium} as follows;

We begin by building an eigenstrain model for the dagger based on \eqref{eq:eigenstrainbasis} and a polynomial basis of the form
\[
\Bigg\{\begin{bmatrix} \varphi_{ij}^1 & 0 & 0\\ 0 & 0 & 0\\0 & 0 &0  \end{bmatrix},\begin{bmatrix} 0 & 0 & 0\\ 0 & \varphi_{ij}^2 & 0\\0 & 0 &0  \end{bmatrix},\begin{bmatrix} 0 & 0 & 0\\ 0 & 0 & 0\\0 & 0 &\varphi_{ij}^3  \end{bmatrix}\Bigg\} \quad \text{for} \quad i,j\in[0,N]
\]
where each of the three scalar diagonal components is a two-dimensional function of the form
\[
\varphi_{ij}(x_1,x_2)=x_1^ix_2^j,
\]
and $n_b=3(N+1)^2$.  These were mapped to basis strain fields using a two-dimensional generalised plane-strain finite element model for the blade.  Briefly, this involves solving a plane-strain version of \eqref{eq:BVP} before applying a constant elastic relaxation in the $x_3$ direction to ensure
\[
\int_A \sigma_{33} dA=0,
\]
over the cross-section.  This was implemented in the MATLAB PDE toolbox using 25,000 quadratic triangular elements.  Material properties for bronze with 9.7\% tin were used based on Vegard's law and the observed lattice parameter.

Through this eigenstrain model and the procedure described earlier, optimal coefficients for a polynomial parameterisation of $d_0$ of the form
\[
d_0(x_1,x_2)=\sum_{i,j=0}^N a_{ij}x_1^ix_2^j,
\]
were found using the Levenberg-Marquardt algorithm within the MATLAB optimisation toolbox.  

The results of this process for $N=4$ and a small amount of regularisation aimed at penalising high-order eigenstrains are shown in Figure \ref{fig:DaggerData}.  Figure \ref{fig:DaggerData}b shows the stress distribution from the least-squares best fit eigenstrain model, \ref{fig:DaggerData}c shows the corresponding predicted distribution for $d_0$ in the same colour scale as the $d$-spacing measurements, and \ref{fig:DaggerData}d shows a histogram of the $d$-spacing residuals compared to the typical peak-fit uncertainty as a reference. 

From the residuals it is clear that this model captures the overall behaviour of the data quite well.  The standard deviation of the residuals was only $1.5\times 10^{-4}\AA$; slightly larger than the typical peak-fit uncertainty of $1\times 10^{-4}\AA$, but not substantially so.  

From the model, the predicted variation in $d_0$ over the data points was of the order of $\pm 1.9\times 10^{-4}\AA$ with an average of $1.10774\AA$.  This is a relatively small variation, but enough to make a significant difference to the predicted stress.  The distribution indicates slightly larger $d_0$ near the surfaces compared to the core of the blade.

Overall, the fitted stress distribution in Figure \ref{fig:DaggerData}b is approximately symmetrical with a core of tension and compressive stresses near the outer surface.  The exception to this rule is the right-most blade edge that experiences very little stress.  Outside of this exception, the data is consistent with casting/quenching processes, potentially with some cold-working (forging) of the blade edge, however this is where our speculation will end.

Once again, does this prove there is $d_0$ variation within the blade according to Figure \ref{fig:DaggerData}c?  Probably not, and perhaps we will never know.  Regardless, we have demonstrated that we are at least able to make predictions from the solid foundation of mechanical equilibrium.

\section{Conclusions}

$d_0$ is often introduced as a simple unknown constant required to convert a measured lattice spacing into elastic strain. In practice, it is neither necessarily constant nor directly accessible, but must instead be inferred within the same physical framework used to interpret the diffraction measurements. The fundamental difficulty is that diffraction measurements alone cannot distinguish between a change in $d_0$ and a hydrostatic elastic strain; the two quantities are observationally indistinguishable and some additional information is always required.

The conventional solution is direct measurement from equivalent stress-relieved material, and this remains the preferred approach where an appropriate coupon can be produced without altering the relevant composition, phase or microstructural state. Where this is not possible, mechanical equilibrium can provide an alternative source of information. At the simplest level, integral force balances can determine a constant $d_0$ while boundary conditions can provide local estimates. More generally, point-wise equilibrium can constrain spatially varying distributions as long the assumption of isotropy is appropriate.

We have developed this idea for several practically important measurement geometries. For one-dimensional axisymmetric systems, equilibrium leads to an explicit first-order differential equation whose solution uniquely determines $d_0(r)$ from radial, hoop and axial lattice-spacings. For measurements on symmetry planes and two-dimensional systems (i.e. plane-stress or plane-strain), an eigenstrain representation provides a convenient regularisation for reconstructing physically admissible stress fields and $d_0$ simultaneously. The accompanying linearised analysis shows that the equilibrium constraint is sufficient to make the reconstruction unique on a symmetry plane from three orthogonal lattice-spacing fields, and more generally in three dimensions from four suitably chosen measurement directions.

The examples we have considered illustrate both the value and the limitations of this approach. For the additively manufactured Inconel cube, equilibrium-based estimates agree closely with direct measurements from a stress-relieved comb. For the Roman medical probe and Persian dagger, where destructive reference measurements are inappropriate, equilibrium permits internally consistent estimates of spatially varying $d_0$ and residual stress. These reconstructions do not, by themselves, prove that the inferred $d_0$ represent reality. However, they do replace an arbitrary reference-spacing assumption with one constrained by the fundamental requirement of mechanical equilibrium.

The central practical message is therefore simple: $d_0$ should not just be treated as an independent calibration constant to be measured when the mechanics of the sample and its symmetry can provide enough information for its reconstruction. Force balance, boundary conditions and point-wise equilibrium form a hierarchy of increasingly powerful constraints that can supplement, test and, in some cases, replace direct stress-free reference measurements -- particularly in situations where destructive approaches are completely inappropriate.

\section{Acknowledgements}

Data presented in this paper was measured using the KOWARI Residual Strain Diffractometer and DINGO Neutron Imaging instrument at the Australian Centre for Neutron Scattering (ACNS) within the Australian Nuclear Science and Technology Organisation (ANSTO).  Access to these instruments was provided through the ACNS User Access Scheme through proposals PP6050 P14244 and P18396 (Inconel cube), P16860, P20100 and P25059 (Roman bronze medical tools) and P21314 (bronze dagger).  In addition to this access, ANSTO provided accommodation and some travel support over the duration of these experiments.  

This research was supported by an AINSE Ltd. Pathway Scholarship awarded to G.N. Parkes.

The Roman medical tools and Persian dagger are owned by the RD Milns Antiquities Museum at the University of Queensland, Australia.  The authors would like to acknowledge the assistance of James Donaldson from the museum for his help in gaining access to these items.

We also acknowledge the Additive Manufacturing Research Laboratory (AMRL) at RISE IVF in Sweden for manufacturing the Inconel cube.

\appendix

\section{Uniqueness of $d_0$ reconstructions on symmetric planes} \label{Sec:ApndxA}

We wish to show that, by including the equilibrium constraint, any reconstruction of $d_0$ from a set of diffraction measurements over a symmetric plane within a sample is unique.  In this endeavour we work with an approximation where we examine small perturbations from an estimated $d_0^*$ of the form
\[
d_0(x)=d_0^*+\delta d_0(x)
\]
where $|\delta d_0|\ll d_0^*$.  As discussed earlier, assuming $\epsilon$ is small, it is conceivable that we can estimate $d_0^*$ to within a fraction of a percent of the true value.

Without loss of generality, we assume the plane in question is orthogonal to the $e_1$ axis, and we also assume we have full knowledge of $d$-spacings in the $\kappa=e_1,e_2$, and $e_3$ directions over the plane.  We also work with the assumption that the material is homogeneous and isotropic.

Symmetry about the plane suggests that all derivatives with respect to $x_1$ are zero, and that $\sigma_{12}=\sigma_{13}=0$ and $\epsilon_{12}=\epsilon_{13}=0$ within the plane.  Working in the Fourier domain, the remaining equations from \eqref{eq:Equilbrium} become
\begin{equation}\label{eq:EquilibriumOnPlane}
\begin{split}
    y_2\hat\sigma_{22}+y_3\hat\sigma_{23}&=0\\
    y_2\hat\sigma_{23}+y_3\hat\sigma_{33}&=0
\end{split}
\end{equation}
where the notation $\hat f(y)=\mathcal{F}(f(x))$ refers to the two-dimensional Fourier transform of $f$ defined for the spatial frequency $\begin{bmatrix}y_2 & y_3\end{bmatrix}^T\in\mathbb{R}^2$.  In this respect it is useful to consider $\sigma$ to be \emph{extended by zero} outside of the sample; in this case, our zero-traction boundary conditions ensure that the equilibrium equations above are valid (in the weak-sense) over the entire plane \cite{wensrich_general_2025}.  In this extended version, all of our unknowns (i.e. $\sigma,\epsilon$ and $\delta d_0$) are only non-zero within the sample (i.e. they are \emph{compactly supported} within $\Omega$), and \eqref{eq:EquilibriumOnPlane} can be considered to apply everywhere on the plane.

Through Hooke's law and a similar argument behind \eqref{eq:fitd0} we can then generate the following set of equations for our unknown strains and $\delta d_0$ in the Fourier domain
\begin{equation}\label{eq:d0_dist}
\begin{bmatrix}
1 & 0 &0  & 
0 &1/d_0^*\\
0 & 1 &0  & 
0 &1/d_0^*\\
0 & 0 &1  & 
0 &1/d_0^*\\
\alpha y_2 & (1+\alpha)y_2 & \alpha y_2  & y_3 & 0 \\
\alpha y_3 & \alpha y_3 & (1+\alpha)y_3  & y_2 & 0 \\
\end{bmatrix}
\begin{bmatrix}
\hat\epsilon_{11}\\
\hat\epsilon_{22}\\
\hat\epsilon_{33}\\
\hat\epsilon_{23}\\
\hat{\delta d_0}
\end{bmatrix}=
\frac{1}{d_0^*}\begin{bmatrix}
    \hat d_1-d_0^*\\
    \hat d_2-d_0^*\\
    \hat d_3-d_0^*\\
    0\\
    0
\end{bmatrix}.
\end{equation}

Note the subtle change in the meaning of this set of equations (as opposed to \eqref{eq:fitd0}).  We are no longer evaluating the unknown strains at a given point; \eqref{eq:d0_dist} defines a set of differential equations for evaluating $\epsilon$ along with $\delta d_0$ over the entire plane when extended to infinity.  Hence if we can solve Equation \eqref{eq:d0_dist}, we would have the complete Fourier transform of the unknowns which can then be recovered by Fourier inversion.

The determinant of the matrix in \eqref{eq:d0_dist} is
\[
\frac{3\alpha+1}{d_0^*}(y_2+y_3)(y_2-y_3),
\]
which is non-zero everywhere in $\mathbb{R}^2$ except for two characteristic lines defined by $y_2=\pm y_3$ (note that $3\alpha+1\ne0$ for all physical materials).  Outside of these characteristics the inverse exists and hence $\hat\epsilon$ and $\hat{\delta d_0}$ are uniquely defined \emph{almost everywhere} in the Fourier domain.  

On the characteristics the inverse does not exist, but we can rely on the well-known Paley-Wiener theorem which states that any function that has compact support in the real domain will be smooth and analytic everywhere in the Fourier domain.  Given that the characteristic planes have measure-zero (i.e. no interior), we can uniquely define our solution within them through analytic continuation (i.e. Taylor series) from outside.  

Through this process (i.e. direct inversion or analytic continuation) we can uniquely define our solution everywhere in the Fourier domain and hence they are uniquely defined everywhere in the real domain as well.

Note that this process is probably not a particularly practical approach to reconstructing $d_0$ as the vanishing determinant near the characteristic planes is likely to introduce numerical instability.  But this is not the point; through this argument we have demonstrated the uniqueness of an accurate first-order approximation of the solution (good to less than a fraction of a percent).  Provided this solution can be found through some stable numerical process, we can be assured that it will be unique.

This was done using the bare minimum of measured data (i.e. $d$-spacings in only three coordinate directions relative to the plane).  In reality, we may have much more information at our disposal and any numerical scheme aiming to simultaneously reconstruct $\epsilon$ and $d_0$ should use all of this data to obtain the best estimate.

We should also note that we do not address the existence of a solution in the above argument. Instead, we assume that existence is guaranteed given the measured data arise from a physically admissible strain (i.e. satisfying equilibrium) and $d_0$ distribution.  Strictly speaking, in the presence of measurement noise this may not be mathematically true; however our assumption is that through the numerical process we employ (e.g. `least squares') our estimate of the solution will be close to the unique solution corresponding to the existing physical reality.

\section{General uniqueness of $d_0$ reconstructions through equilibrium}\label{Sec:ApndxB}

Further to \ref{Sec:ApndxA}, we can use a similar approach to show a more general result implying that the constraint of equilibrium allows the unique reconstruction of $d_0$ from a sufficient set of measured $d$-spacings.  The argument is as follows;

\begin{figure}
    \centering
    \includegraphics[width=0.5\linewidth]{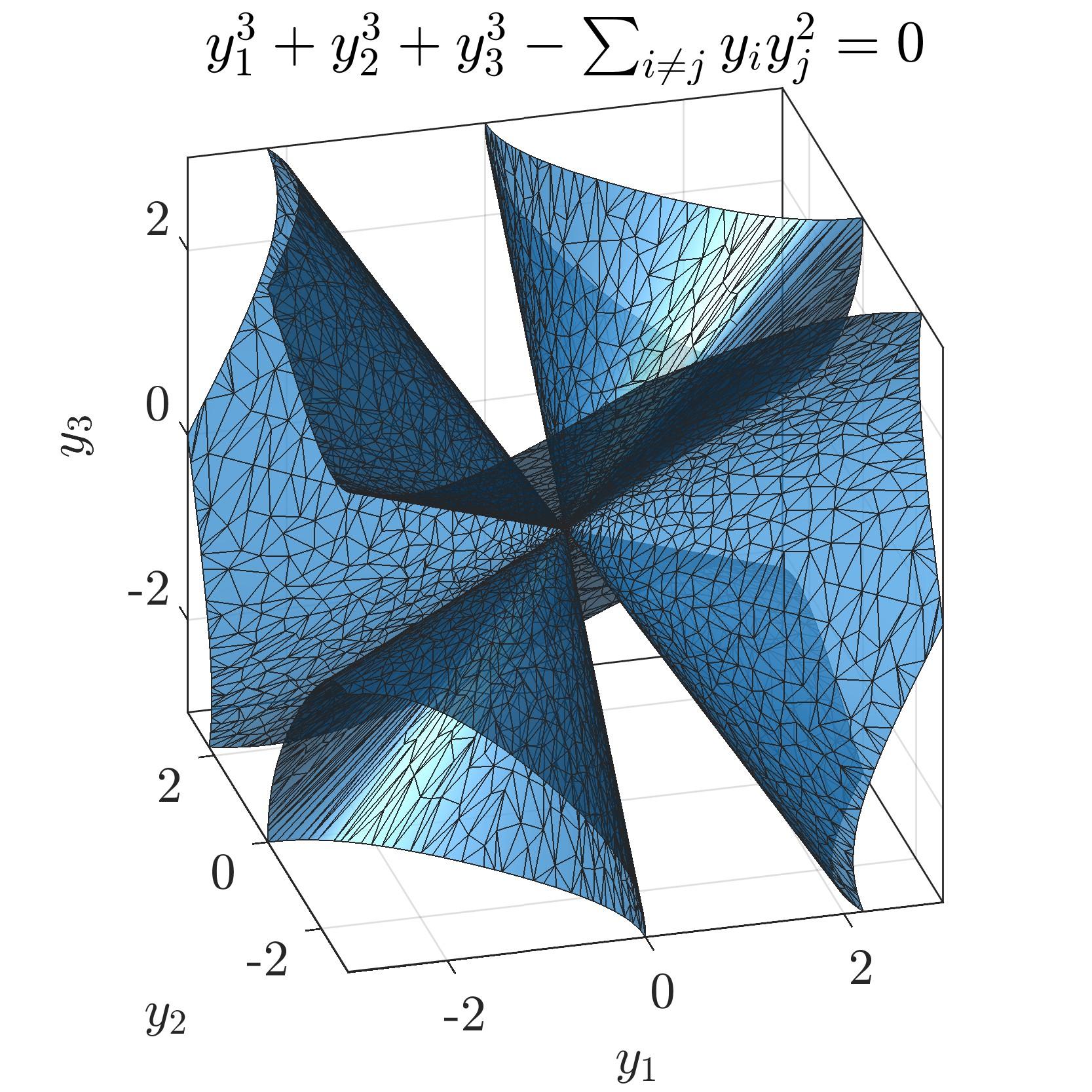}
    \caption{Characteristic surfaces of \eqref{eq:d0_dist_full}}
    \label{fig:PolyRoots}
\end{figure}

Say we have a set of measured $d$-spacings in four different directions at every point within a sample.  Once again, this sample is finite and so the strain and $d_0$ distribution are both compactly supported in $\mathbb{R}^3$.  For convenience, the directions we choose are $\kappa=\{e_1,e_2,e_3,e_s\}$, where $e_s$ is the symmetric direction
\[
e_s=(e_1+e_2+e_3)/\sqrt3.
\]

As in \ref{Sec:ApndxA}, we also assume variation in $d_0$ is small and linearise around an approximation of the form
\[
d_0(x)=d_0^*+\delta d_0(x).
\]
Assembling equations from the measurements along with equilibrium we build the system in the Fourier domain
\begin{equation}\label{eq:d0_dist_full}
\begin{bmatrix}
1 & 0 &0  & 0 & 0& 0 & 1/d_0^*\\
0 & 1 &0  & 0 & 0& 0 & 1/d_0^*\\
0 & 0 &1  & 0 & 0& 0 & 1/d_0^*\\
1/3 & 1/3 & 1/3 & 2/3 & 2/3 &2/3 & 1/d_0^*\\
(1+\alpha) y_1 & \alpha y_1 & \alpha y_1  & 0 & y_3 & y_2 & 0\\
\alpha y_2 & (1+\alpha)y_2 & \alpha y_2  & y_3 & 0 &y_1 & 0\\
\alpha y_3 & \alpha y_3 & (1+\alpha)y_3  & y_2& y_1 & 0 & 0 \\
\end{bmatrix}
\begin{bmatrix}
\hat\epsilon_{11}\\
\hat\epsilon_{22}\\
\hat\epsilon_{33}\\
\hat\epsilon_{23}\\
\hat\epsilon_{13}\\
\hat\epsilon_{12}\\
\hat{\delta d_0}
\end{bmatrix}=
\frac{1}{d_0^*}\begin{bmatrix}
    \hat d_1-d_0^*\\
    \hat d_2-d_0^*\\
    \hat d_3-d_0^*\\
    \hat d_s-d_0^*\\
    0\\
    0\\
    0
\end{bmatrix},
\end{equation}
where the measurement $d_s$ refers to the $d$-spacing in the symmetric direction.

The determinant of this system can be written
\[
\frac{2}{3d_0^*}(3\alpha+1)\Big(y_1^3+y_2^3+y_3^3-\sum_{i\ne j}y_iy_j^2\Big),
\]
and while the zeros of this polynomial are much more complex than the symmetric case in \ref{Sec:ApndxA}, they still define a set of zero-measure surfaces in $\mathbb{R}^3$ (see Figure \ref{fig:PolyRoots}).  Outside of these surfaces (i.e. almost everywhere in $\mathbb{R}^3$), \eqref{eq:d0_dist_full} is invertible and uniquely defines our solution.

As before, Paley-Wiener guarantees our compactly supported functions $\epsilon(x)$ and $\delta d_0(x)$ transform to smooth and analytic functions $\hat\epsilon(y)$ and $\hat{\delta d_0}(y)$ in the Fourier domain.  So by analytic continuation, we can again uniquely define our unknowns everywhere in the Fourier domain and hence $\epsilon$ and $\delta d_0$ are uniquely defined everywhere within the sample through Fourier inversion.

\bibliography{references}

@article{maekawa_stress-free_2012,
	title = {Stress-free reference for neutron diffraction measurement of residual stress in butt-welded joints of austenitic stainless steel pipes},
	volume = {6},
	number = {9},
	journal = {Journal of Solid Mechanics and Materials Engineering},
	author = {Maekawa, Akira and Takahashi, Tsuneo and Tsuji, Takashi and Suzuki, Hiroshi and Moriai, Atsushi},
	year = {2012},
	pages = {950--964},
}

@article{rogante_stress-free_2000,
	title = {The stress-free reference sample: the problem of the determination of the interplanar distance d0},
	volume = {276},
	journal = {Physica B: Condensed Matter},
	author = {Rogante, Massimo},
	year = {2000},
	pages = {202--203},
}

@article{kesavan_nair_residual_1995,
	title = {Residual stresses of types {II} and {III} and their estimation},
	volume = {20},
	number = {1},
	journal = {Sadhana},
	author = {Kesavan Nair, P and Vasudevan, R},
	year = {1995},
	pages = {39--52},
}

@article{withers_residual_2001,
	title = {Residual stress. {Part} 2–{Nature} and origins},
	volume = {17},
	number = {4},
	journal = {Materials science and technology},
	author = {Withers, Philip J and Bhadeshia, HKDH},
	year = {2001},
	pages = {366--375},
}

@article{hauk_structural_1997,
	title = {Structural and residual stress analysis by nondestructive methods: {Evaluation}-{Application}-{Assessment}},
	author = {Hauk, Viktor},
	year = {1997},
}

@book{fitzpatrick_analysis_2003,
	title = {Analysis of residual stress by diffraction using neutron and synchrotron radiation},
	publisher = {CRC Press},
	author = {Fitzpatrick, Michael E and Lodini, Alain},
	year = {2003},
}

@article{gnaupel-herold_measurement_2005,
	title = {Measurement of residual stress in materials using neutrons {Proceedings} of a technical meeting held in {Vienna} 13-17 {October}, 2003},
	journal = {International Atomic Energy Agency,(June)},
	author = {Gnäupel-Herold, T and Schneider, R and Mikula, P and Paranjpe, SK and Teixeira, J and Torok, G and Venter, A and Youtsos, AG},
	year = {2005},
	pages = {1--99},
}

@book{noyan_residual_2013,
	title = {Residual stress: measurement by diffraction and interpretation},
	publisher = {Springer},
	author = {Noyan, Ismail C and Cohen, Jerome B},
	year = {2013},
}

@book{kisi_applications_2012,
	title = {Applications of neutron powder diffraction},
	volume = {15},
	publisher = {Oxford University Press},
	author = {Kisi, Erich H and Howard, Christopher J},
	year = {2012},
}

@article{bhadeshia_stress_1991,
	title = {Stress induced transformation to bainite in {Fe}–{Cr}–{Mo}–{C} pressure vessel steel},
	volume = {7},
	issn = {0267-0836, 1743-2847},
	url = {https://journals.sagepub.com/doi/full/10.1179/mst.1991.7.8.686},
	doi = {10.1179/mst.1991.7.8.686},
	language = {en},
	number = {8},
	urldate = {2026-06-08},
	journal = {Materials Science and Technology},
	author = {Bhadeshia, H. K. D. H. and David, S. A. and Vitek, J. M. and Reed, R. W.},
	month = aug,
	year = {1991},
	pages = {686--698},
}

@article{withers_methods_2007,
	title = {Methods for obtaining the strain-free lattice parameter when using diffraction to determine residual stress},
	volume = {40},
	issn = {0021-8898},
	url = {https://journals.iucr.org/paper?S0021889807030269},
	doi = {10.1107/S0021889807030269},
	number = {5},
	urldate = {2026-06-08},
	journal = {Journal of Applied Crystallography},
	author = {Withers, P. J. and Preuss, M. and Steuwer, A. and Pang, J. W. L.},
	month = oct,
	year = {2007},
	pages = {891--904},
}

@article{youtsos_vamas_2000,
	title = {{VAMAS} {TWA} 20 workshops held at {ORNL} and {JRC}-{Petten}},
	volume = {11},
	issn = {1044-8632, 1931-7352},
	url = {http://www.tandfonline.com/doi/abs/10.1080/10448630008233718},
	doi = {10.1080/10448630008233718},
	language = {en},
	number = {2},
	urldate = {2026-04-16},
	journal = {Neutron News},
	author = {Youtsos, Tassos and Ohms, Carsten and Hubbard, Camden},
	month = jan,
	year = {2000},
	pages = {2--4},
}

@article{jahed_axisymmetric_1997,
	title = {An {Axisymmetric} {Method} of {Elastic}-{Plastic} {Analysis} {Capable} of {Predicting} {Residual} {Stress} {Field}},
	volume = {119},
	issn = {0094-9930, 1528-8978},
	url = {https://asmedigitalcollection.asme.org/pressurevesseltech/article/119/3/264/454416/An-Axisymmetric-Method-of-ElasticPlastic-Analysis},
	doi = {10.1115/1.2842303},
	language = {en},
	number = {3},
	urldate = {2026-04-16},
	journal = {Journal of Pressure Vessel Technology},
	author = {Jahed, H. and Dubey, R. N.},
	month = aug,
	year = {1997},
	pages = {264--273},
}

@article{daymond_analysis_2002,
	title = {Analysis of neutron diffraction strain measurement data from a round robin sample},
	volume = {37},
	number = {1},
	journal = {The Journal of Strain Analysis for Engineering Design},
	author = {Daymond, MR and Johnson, MW and Sivia, DS},
	year = {2002},
	pages = {73--85},
}

@article{wensrich_measurement_2012,
	title = {Measurement and analysis of the stress distribution during die compaction using neutron diffraction},
	volume = {14},
	copyright = {http://www.springer.com/tdm},
	issn = {1434-5021, 1434-7636},
	url = {http://link.springer.com/10.1007/s10035-012-0366-8},
	doi = {10.1007/s10035-012-0366-8},
	language = {en},
	number = {6},
	urldate = {2026-04-16},
	journal = {Granular Matter},
	author = {Wensrich, C. M. and Kisi, E. H. and Zhang, J. F. and Kirstein, O.},
	month = nov,
	year = {2012},
	pages = {671--680},
}

@article{abbey_reconstruction_2012,
	title = {Reconstruction of axisymmetric strain distributions via neutron strain tomography},
	volume = {270},
	journal = {Nuclear Instruments and Methods in Physics Research Section B: Beam Interactions with Materials and Atoms},
	author = {Abbey, Brian and Zhang, Shu Yan and Vorster, Wim and Korsunsky, Alexander M},
	year = {2012},
	pages = {28--35},
}

@article{stacey_measurement_1985,
	title = {Measurement of residual stresses by neutron diffraction},
	volume = {20},
	copyright = {https://journals.sagepub.com/page/policies/text-and-data-mining-license},
	issn = {0309-3247, 2041-3130},
	url = {https://journals.sagepub.com/doi/10.1243/03093247V202093},
	doi = {10.1243/03093247V202093},
	language = {en},
	number = {2},
	urldate = {2026-04-16},
	journal = {The Journal of Strain Analysis for Engineering Design},
	author = {Stacey, A and MacGillivary, H J and Webster, G A and Webster, P J and Ziebeck, K R A},
	month = apr,
	year = {1985},
	pages = {93--100},
}

@article{wensrich_uniqueness_2026,
	title = {Uniqueness of solutions in high-energy x-ray based ‘eigenstrain tomography’ and other inverse eigenstrain problems: {Counter} examples and necessary conditions for well-posedness},
	volume = {212},
	issn = {00225096},
	shorttitle = {Uniqueness of solutions in high-energy x-ray based ‘eigenstrain tomography’ and other inverse eigenstrain problems},
	url = {https://linkinghub.elsevier.com/retrieve/pii/S0022509626000967},
	doi = {10.1016/j.jmps.2026.106596},
	language = {en},
	urldate = {2026-04-02},
	journal = {Journal of the Mechanics and Physics of Solids},
	author = {Wensrich, C.M. and Holman, S. and Lionheart, W.B.L. and Courdurier, M. and Jackson, R.R.},
	month = jun,
	year = {2026},
	pages = {106596},
}

@article{uzun_eigenstrain_2025,
	title = {Eigenstrain {Tomography}: {Full}-{Field} {Residual} {Stress} {Reconstruction} via {Polycrystalline} {Diffraction} {Projections}},
	journal = {Acta Materialia},
	author = {Uzun, Fatih and Korsunsky, Alexander M},
	year = {2025},
	pages = {121872},
}

@article{ziegler_eigenstrain_2004,
	title = {Eigenstrain controlled deformation and stress states},
	volume = {23},
	number = {1},
	journal = {European Journal of Mechanics-A/Solids},
	author = {Ziegler, Franz},
	year = {2004},
	pages = {1--13},
}

@article{irschik_eigenstrain_2001,
	title = {Eigenstrain without stress and static shape control of structures},
	volume = {39},
	number = {10},
	journal = {AIAA journal},
	author = {Irschik, Hans and Ziegler, Franz},
	year = {2001},
	pages = {1985--1990},
}

@article{wensrich_general_2025,
	title = {General reconstruction of elastic strain fields from their {Longitudinal} {Ray} {Transform}},
	volume = {85},
	number = {2},
	journal = {SIAM Journal on Applied Mathematics},
	author = {Wensrich, Chris and Holman, Sean and Lionheart, William and Courdurier, Matias and Polyakova, Anna and Svetov, Ivan and Doubikin, Ty},
	year = {2025},
	pages = {945--960},
}

@article{uzun_voxel-based_2023,
	title = {Voxel-{Based} {Full}-{Field} {Eigenstrain} {Reconstruction} of {Residual} {Stresses}},
	volume = {25},
	number = {14},
	journal = {Advanced Engineering Materials},
	author = {Uzun, Fatih and Korsunsky, Alexander M},
	year = {2023},
	pages = {2201502},
}

@article{korsunsky_variational_2007,
	title = {Variational eigenstrain analysis of residual stresses in a welded plate},
	volume = {44},
	number = {13},
	journal = {International Journal of Solids and Structures},
	author = {Korsunsky, Alexander M and Regino, Gabriel M and Nowell, David},
	year = {2007},
	pages = {4574--4591},
}

@book{mura_micromechanics_2013,
	title = {Micromechanics of defects in solids},
	publisher = {Springer Science \& Business Media},
	author = {Mura, Toshio},
	year = {2013},
}

@article{korsunsky_eigenstrain_2008,
	title = {Eigenstrain analysis of residual strains and stresses},
	volume = {41},
	number = {1},
	journal = {The Journal of Strain Analysis for Engineering Design},
	author = {Korsunsky, Alexander M.},
	year = {2008},
	pages = {29--43},
}

@book{korsunsky_teaching_2017,
	title = {A teaching essay on residual stresses and eigenstrains},
	publisher = {Butterworth-Heinemann},
	author = {Korsunsky, Alexander M},
	year = {2017},
}

@article{uzun_tomographic_2024,
	title = {Tomographic eigenstrain reconstruction for full-field residual stress analysis in large scale additive manufacturing parts},
	volume = {81},
	issn = {22148604},
	url = {https://linkinghub.elsevier.com/retrieve/pii/S2214860424000733},
	doi = {10.1016/j.addma.2024.104027},
	language = {en},
	urldate = {2025-08-18},
	journal = {Additive Manufacturing},
	author = {Uzun, Fatih and Basoalto, Hector and Liogas, Konstantinos and Slim, Mohamed Fares and Lee, Tung Lik and Besnard, Cyril and Wang, Zifan Ivan and Chen, Jingwei and Dolbnya, Igor P. and Korsunsky, Alexander M.},
	month = feb,
	year = {2024},
	pages = {104027},
}

@article{wensrich_residual_2024,
	title = {Residual stress in additively manufactured {Inconel} cubes; {Selective} {Laser} {Melting} versus {Electron} {Beam} {Melting} and a comparison of modelling techniques},
	volume = {244},
	issn = {02641275},
	url = {https://linkinghub.elsevier.com/retrieve/pii/S0264127524004829},
	doi = {10.1016/j.matdes.2024.113108},
	language = {en},
	urldate = {2025-08-16},
	journal = {Materials \& Design},
	author = {Wensrich, C.M. and Luzin, V. and Hendriks, J.N. and Pant, P. and Gregg, A.W.T.},
	month = aug,
	year = {2024},
	pages = {113108},
}

@article{wensrich_well-posedness_2025,
	title = {Well-posedness and trivial solutions to inverse eigenstrain problems},
	volume = {321},
	issn = {00207683},
	url = {https://linkinghub.elsevier.com/retrieve/pii/S0020768325002914},
	doi = {10.1016/j.ijsolstr.2025.113505},
	language = {en},
	urldate = {2025-08-16},
	journal = {International Journal of Solids and Structures},
	author = {Wensrich, C.M. and Holman, S. and Lionheart, W.B.L. and Luzin, V. and Cuskelly, D. and Kirstein, O. and Salvemini, F.},
	month = oct,
	year = {2025},
	pages = {113505},
}

\end{document}